\documentclass[iop,preprint2]{emulateapj1}
\usepackage{ulem}
\usepackage{graphicx}
\usepackage{apjfonts}
\usepackage{rotating}
\usepackage[USenglish]{babel}
\usepackage{color}

\newcommand{\wcen}{$\omega$~Cen}
\usepackage{longtable}

\begin{document}

\title{THE STATE-OF-THE-ART \textit{HST} ASTRO-PHOTOMETRIC ANALYSIS OF
  THE CORE OF $\omega$~CENTAURI. III. THE MAIN SEQUENCE's
  MULTIPLE POPULATIONS GALORE$^{\ast}$}

\author{A.\ Bellini\altaffilmark{1},
A.\ P.\ Milone\altaffilmark{2},
J.\ Anderson\altaffilmark{1},
A.\ F.\ Marino\altaffilmark{2},
G.\ Piotto\altaffilmark{3,4},
R.\ P.\ van der Marel\altaffilmark{1},
L.\ R.\ Bedin\altaffilmark{3}, and
I.\ R.\ King\altaffilmark{5}}

\altaffiltext{1}{Space Telescope Science Institute, 3700 San Martin
  Dr., Baltimore, MD 21218, USA}
\altaffiltext{2}{Research School of Astronomy \& Astrophysics,
  Australian National University, Mt Stromlo Observatory, via Cotter
  Rd, Weston, ACT 2611, Australia}
\altaffiltext{3}{Istituto Nazionale di Astrofisica, Osservatorio
  Astronomico di Padova, v.co dell'Osservatorio 5, Padova, I-35122,
  Italy}
\altaffiltext{4}{Dipartimento di Fisica e Astronomia ``Galileo
  Galilei'', Universit\`a di Padova, Vicolo dell'Osservatorio 3,
  Padova I-35122, Italy}
\altaffiltext{5}{Department of Astronomy, University of Washington,
  Box 351580, Seattle, 98195, WA, USA}
\altaffiltext{$^\ast$}{Based on archival observations with the
  NASA/ESA \textit{Hubble Space Telescope}, obtained at the Space
  Telescope Science Institute, which is operated by AURA, Inc., under
  NASA contract NAS 5-26555.}
\received{May 22, 2017}
\revised{June 20, 2017}
\accepted{June 21, 2017}

\email{bellini@stsci.edu}

\begin{abstract} {
We take advantage of the exquisite quality of the \textit{Hubble Space
  Telescope} 26-filter astro-photometric catalog of the core of
$\omega$~Cen presented in the first paper of this series and the
empirical differential-reddening correction presented in the second
paper in order to distill the main sequence into its constituent
populations.  To this end, we restrict ourselves to the five most
useful filters: the magic ``trio'' of F275W, F336W, and F438W, along
with F606W and F814W.  We develop a strategy for identifying color
systems where different populations stand out most distinctly, then we
isolate those populations and examine them in other filters where
their sub-populations also come to light.  In this way, we have
identified at least 15 sub-populations, each of which has a
distinctive fiducial curve through our 5-dimensional photometric
space.  We confirm the MSa to be split into two subcomponents, and
find that both the bMS and the rMS are split into three
subcomponents. Moreover, we have discovered two additional MS
groups:\ the MSd (which has three subcomponents) shares similar
properties with the bMS, and the MSe (which has four subcomponents),
has properties more similar to those of the rMS.  We examine the
fiducial curves together and use synthetic spectra to infer relative
heavy-element, light-element, and Helium abundances for the
populations.  Our findings show that the stellar populations and star
formation history of \wcen\ are even more complex than inferred
previously.  Finally, we provide as a supplement to the original
catalog a list that identifies for each star which population it most
likely is associated with.}
\end{abstract}


\maketitle

\section{Introduction}
\label{sec:intro}

Omega Centauri (\wcen) was long believed to be a ``missing link''
transition object between globular clusters (GCs) and dwarf
spheroidals (see, e.g., \citealt{2003MNRAS.346L..11B}).  Indeed, in
August of 2001 a conference was convened in Cambridge, England, with
the expressed goal of debating which bin to place it in
(\citealt{2002ASPC..265.....V}).

It had been known since the seventies (\citealt{1973MNRAS.162..207C})
that $\omega$~Cen's stellar chemistry is complex, with large
variations in C, N, and iron. Therefore, the discovery of multiple red
giant branches (RGBs, \citealt{1999Natur.402...55L,
  2000A&A...362..895H, 2000ApJ...534L..83P}) that could be associated
with different broad metallicity peaks was understandable and easily
accepted by the astronomical community.

At the conference however, Anderson presented new results from his
thesis (\citealt{anderson97, 2002ASPC..265...87A}) based on
photometric techniques optimized for \textit{Hubble Space Telescope}
(\textit{HST}) data that showed that the cluster's main sequence (MS)
was split into two clearly distinct sequences.  There were several
reasons that this was a surprising finding.  First, spectroscopic data
suggested broad overlapping distributions, not distinct,
well-separated populations.  Even more curious was the fact that in
the \textit{HST} $m_{\rm F606W}$ vs.\ $m_{\rm F606W}-m_{\rm F814W}$
CMD explored, the reddest MS was the most populous one (see
\citealt{2004ApJ...605L.125B}), while we knew from spectroscopy that
the lower-metallicity stars were the more abundant in the cluster
(\citealt{1995ApJ...447..680N}).

At that time, there was no plausible explanation but to assume a large
difference in helium between the two MSs, as suggested since the
beginning by J.\ Norris, the referee of the
\citet{2004ApJ...605L.125B} paper. Indeed, even now a large He
difference ($\Delta Y \sim 0.14$, \citealt{2012AJ....144....5K})
between the two main components of $\omega$~Cen is the only available
interpretation of the photometric and spectroscopic observational
facts.

The conference ended without a clear binary resolution of \wcen's
dwarf spheroidal or GC nature, which left it as a unique anomaly.  In
the years since, we have come to understand that the non-singular
nature of \wcen's populations are actually just an extreme example of
a multiple-population phenomenon that is present in all clusters.

\textit{HST} has played a crucial role in this research field.
\citet{2015AJ....149...91P} showed that the multiple populations are
present in all the 57 GCs observed within their \textit{HST} Treasury
program and related \textit{HST} ancillary projects, though multiple
populations exhibit different properties in different clusters
(\citealt{2017MNRAS.464.3636M}).  In some cases, the population
complexity is quite intriguing, as in M2 (NGC~7089,
\citealt{2015MNRAS.447..927M}) or NGC~2808
(\citealt{2007ApJ...661L..53P, 2015ApJ...808...51M,
  2015ApJ...810L..13B}).  However, none of these clusters reaches the
incredible multiplicity of $\omega$~Cen.  Indeed, this paper will show
that $\omega$~Cen is even more complicated than we thought.

The results presented here are the product of a massive effort, and
represent a continuation of what we published in
\citet{2010AJ....140..631B}. Paper~I of this series (\citealt{b+17a})
describes the photometric techniques we adopted and applied to 650
individual exposures in 26 different bands.  The photometry has been
corrected for differential reddening and zero-point spatial variations
in \citet[Paper~II]{b+17b}.  In this paper, we analyze the CMDs and
the so-called ``chromosome'' maps (\citealt{2017MNRAS.464.3636M}) of
the MS of the cluster, and finally identified at least 15 distinct
stellar populations.  It stands to reason that a detailed
understanding of the complex multiple populations in the cluster,
which we will (qualitatively) relate to differences in Y, C, N, O, and
Fe in the final part of the paper, will require a huge interpretative
effort in the years to come.

The paper is organized as follows. In Section~\ref{s:sample} we
present the stellar sample. In Section~\ref{s:mps} we describe the
iterative procedures used to extract the different stellar
populations. Section~\ref{s:ov} presents an overview of the multiple
stellar populations we identified, and Section~\ref{s:ana} is
dedicated to a first qualitative analysis of the color difference
among the 15 populations, and to the implications in terms of
differences in Y, C, N, O, and Fe content.

\section{Sample Selection}
\label{s:sample}

\subsection{Choosing the optimal filter set}
\label{ss:filterset}

The 26 filters of our photometric catalog allow us to construct a very
large number of distinct CMDs. Some CMDs based on specific filters are
able to separate distinct sequences more clearly than others.  Key
molecular absorption bands (OH, NH, CN, CH) fall in the F275W, F336W
and F438W bandpasses, making observations through these filters
particularly sensitive to the fingerprints of light-element abundance
differences (see, e.g., \citealt{2015AJ....149...91P}, their Fig.~1).
The F606W and F814W bandpasses are virtually insensitive to
light-element abundance variations, but are good proxies of
temperature differences (and hence, He and Fe content) among MS stars
of a given brightness.  Photometry in most of the medium- and
narrow-band filters in our catalog is available only for a subset of
stars, because of the smaller field-of-view covered by these
observations, the low signal-to-noise at the MS level, and the smaller
number of available images (see, e.g. Fig.~1 and Table~1 of
Paper~I). In addition, due to crowding, photometry in WFC3/IR filters
is not precise enough at the level of the MS for high-precision
sequence analysis (see discussion in Sect.~3.5 of Paper~I).

For these reasons, we limited our multiple-population selection
procedures here to only five filters:\ the so-called ``magic trio''
(F275W, F336W and F438W), and the two optical bands F606W and F814W.
Exposures taken with these five filters cover the largest FoV in our
dataset, and have the largest number of contributing exposures:\ this
guarantees us to maximize the number of MS stars that can be studied
with the smallest photometric errors.  Finally, a limited set of
filters also helps us to minimize the impact of selection effects.  We
will, however, make use of a broader selection of WFC3/UVIS filters
later in Sect.~5, which is dedicated to a qualitative abundance
analysis of each population based on comparisons with synthetic
spectra.

\subsection{Choosing the best-measured stars}
\label{ss:beststars}

As described in Paper~I, photometry in our catalog is measured through
three different methods.  Method one involves fitting a position and
flux for each star in each exposure; it works best for bright,
unsaturated stars. Method two involves forced photometry using the
inner 3$\times$3 pixels of each source at the average position
transformed into each exposure; it is best suited for relatively faint
stars, while method three uses the brightest 4 pixels and weights them
by the expected values of the PSF in those pixels; it is optimized for
extremely low S/N stars. (We refer the interested reader to Paper~I
for a detailed description of the data-reduction processes.)
Saturation in the five selected filters kicks in just above the base
of the RGB. Here we are focused on the bright part of the MS, where
method-two offers the most-precise measurements. Therefore, method-two
photometry is the one we make use of throughout this paper.

Our photometric catalog, containing over 470,000 sources within the
central $4\farcm3 \times 4\farcm3$, offers several quality parameters
that can be used to sift out poorly-measured stars, namely:\ (i) the
quality-of-fit (\texttt{QFIT}), which discriminates between sources
that are fit well by the PSF and extended sources or blends; (ii) the
photometric rms among multiple independent measurements, (iii) the
local sky-background rms, and (iv) the neighbor-contamination
parameter \textit{o}, which tells us the fraction of the flux due to
neighbors within the PSF fitting radius of of a given star and the
star's flux itself.

To select the best-measured stars, we started from the star list we
used in Paper~II to derive a high-precision, differential-reddening
map of the core of the cluster (see their Sect.~2.2 for details). In a
nutshell, this star list is obtained by removing poorly-measured
sources according to all four of the quality parameters mentioned
above. In addition, the list contains only stars with a measured
proper motion consistent with the cluster's bulk motion. The list used
in Paper~II contains 72,609 well measured, proper-motion-selected
cluster stars.

We showed in Paper~II that differential-reddening in the core of the
cluster can vary by up to $\sim$10\%, with a typical standard
deviation of about 4\%. This is generally not a concern for most
scientific applications, since the most-relevant features on a CMD can
be easily recognized without the need for a differential-reddening
correction. Here we aim at characterizing the finest details in a CMD,
and we must correct our photometry for differential-reddening effects.
To do this, we closely followed the prescriptions given in Paper~II.
To minimize the impact of possible systematic errors related to edge
effects in the differential-reddening correction, we further
restricted our star list to include only those stars that had enough
reddening-reference stars within 600 pixels (or 24$^{\prime\prime}$)
to enable an empirical differential-reddening correction.  This can
easily be done using the information contained in the ``radius map''
extension of the \texttt{fits} file we published in Paper~II. This
restriction limited the area covered by selected stars from about
$4\farcm 3 \times 4\farcm 3$ to the inner $3\farcm 3 \times 3\farcm
3$.

The final catalog contains 69$\,$536 high-photometric-quality stars
measured in all five filters, and extends from about 2 magnitudes
above the turn-off ($m_{\rm F606W}$$\sim$16) down to about 3.5
magnitudes below the turn-off ($m_{\rm F606W}$$\sim$21.5). The
limiting factor at the faint end is represented by the insufficient
signal-to-noise in F275W, while saturation dictates the bright-end
limit.

\begin{figure*}[!t]
\centering
\includegraphics[angle=-90,width=16cm]{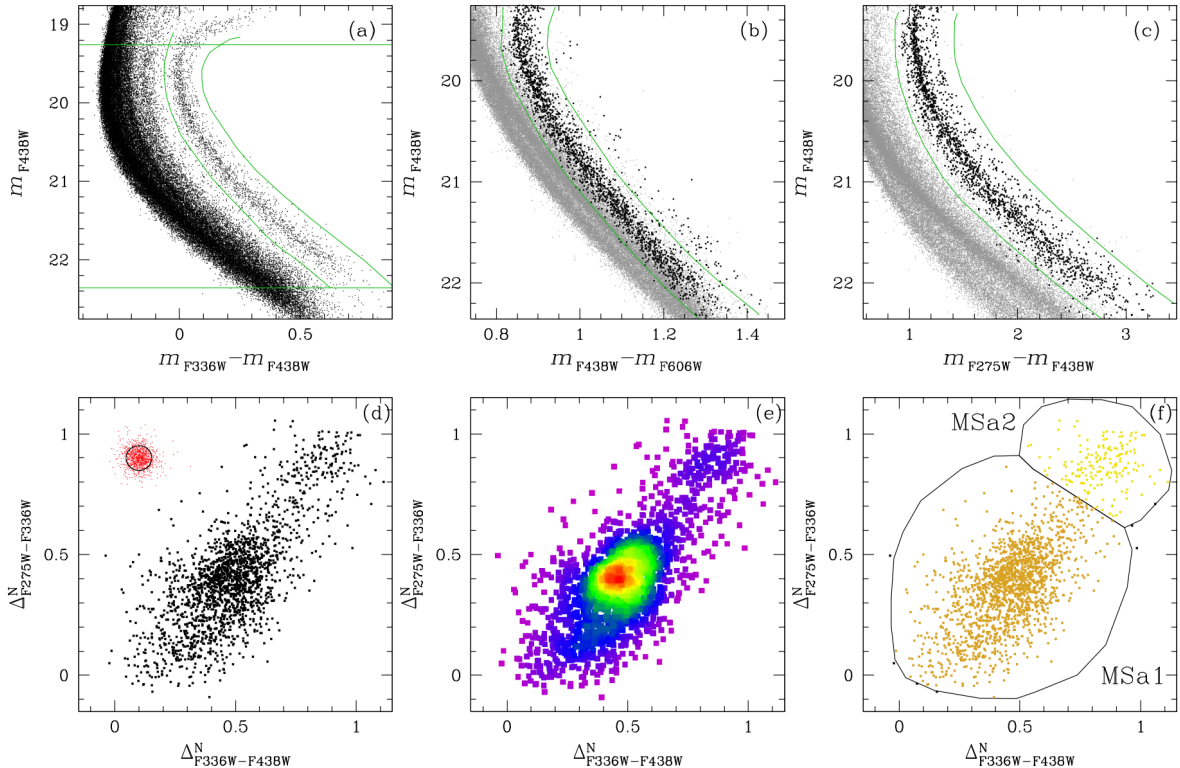}
\vskip -.3cm
\caption{\small{Illustration of the selection procedures we applied to
    isolate MSa stars. (a): Preliminary selection of MSa candidates on
    the $m_{\rm F336W}-m_{\rm F438W}$ CMD (within the green
    lines). (b) and (c):\ selection refinements using two CMDs of
    different color. In black we show MSa stars selected from the
    previous panel, in gray the rest of the MS. Rejected stars are
    those outside the two green lines. In panel (d) we show the
    $\Delta_{m_{\rm F275W}-m_{\rm F336W}}^{\rm N}$
    vs.\ $\Delta_{m_{\rm F336W}-m_{\rm F438W}}^{\rm N}$ chromosome map
    of MSa stars that survived our selections (a)+(b)+(c), in
    black. Red points show the distribution of a single sequence as
    predicted by only photometric and
    differential-reddening-correction errors. The Hess diagram of the
    chromosome map is in panel (e). It is clear that there is a main
    component around (0.45,0,45) in the Hess diagram, which elongated
    shape toward (0,0) might indicate additional substructures, and a
    secondary component at around (0.9,0.9). We defined two MSa
    subpopulations in Panel (f), corresponding to the two main clumps
    in the chromosome map:\ MSa1 (in dark yellow) and MSa2 (in light
    yellow).\\~\\}}
\label{f:msa}
\end{figure*}

\section{The Main Sequence unfolds}
\label{s:mps} 

The MSs of the different stellar populations in \wcen\ typically
overlap each other at different magnitude levels in different CMDs,
and teasing them apart involves a careful, iterative process.  We
adopted the following general approach to identify distinct
populations:\
\begin{itemize}
\item{We start with a preliminary selection of a given population on
  the CMD where this population stands out most clearly (as a spread
  sequence or a bimodal color distribution at fixed magnitude). We
  experimented with all possible combinations of filters to construct
  CMDs, and selected the one that allows the best separation between
  stars of the analyzed population and the rest of the MS. This
  initial, preliminary selection invariably contains contaminants from
  other populations.}
\item{We plot these preliminary-selected stars on a few different CMDs
  in which outliers are most-easily identified and rejected. These
  CMDs are again chosen by trial-and-error as those optimizing the
  identification of outliers.}
\item{We make use of rectified and parallelized two-color diagrams,
  so-called ``chromosome maps'', to highlight finer population
  structures (see \citealt{2015MNRAS.447..927M,2015ApJ...808...51M}).}
\item{Once stars belonging to a given population have been identified
  and selected, we removed them from the list, and repeated the
  process on a different population.}
\item{We repeat the entire process until no more clear distinct
  sequences can be identified.}
\end{itemize}

This iterative procedure of population tagging, refining, selecting
and removing (TRSR) will prove to be quite effective.  Given the
tangled weave intertwined by the MPs on the MS of \wcen, we began by
identifying those populations that clearly stand out on either side of
the MS. Then, we proceeded toward the more intricate central MS
region.

For consistency and simplicity, in the following we will always use
$m_{\rm F438W}$ magnitudes as the $y$ axis of our CMDs, while we let
the CMD color vary. Hence, we will hereafter identify a CMD solely by
its color (unless specifically stated otherwise).

\subsection{The MSa}
\label{ss:msa}

The MSa is clearly isolated from the rest of the MS in the $m_{\rm
  F336W}-m_{\rm F438W}$ CMD (panel a of Fig.~\ref{f:msa}).  We limited
our selection in the magnitude range 19.26$\leq$$m_{\rm
  F438W}$$\leq$22.36. The bright limit is slightly above the MSa
turn-off, while beyond the faint limit the number of available MSa
stars suddenly drops, due to incompleteness.  (Because of the peculiar
isolation of MSa stars from the rest of the MS, we pushed the bright
limit above the turn-off level. The bright limit will be set to a
fainter level, $m_{\rm F438W}$=20.16, for the analysis of the other
populations.)  Magnitude cuts are indicated by the two horizontal
green lines in panel (a). We drew by hand the two lines (also in
green, following the MSa profile) that delimit the blue and the red
boundaries of MSa stars.

We decided to manually define the color boundaries used to delimit the
MSa (as well as all those of all other MSs) because, although an
automatic procedure would be in principle more easily repeatable, no
machine can be as accurate and precise as the human eye for this
task. Our selections are always clearly highlighted at each stage in
all the figures.

Next, preliminary-selected MSa stars are plotted on the $m_{\rm
  F438W}-m_{\rm F606W}$ CMD (in black in panel b).  The remaining,
unidentified MS stars are shown in gray. A few outliers (possibly
binaries and/or stars that do not belong to the MSa population)
clearly stand out, mostly on the red side of the CMD.  We restricted
our MSa selection to only those stars within the two green lines in
panel (b). (For consistency, unless stated otherwise, green lines will
hereafter always indicate our selection boundaries, black points will
mark selected stars from the previous panel, while all other stars
will be in gray.) Panel (c) of Fig.~\ref{f:msa} shows the $m_{\rm
  F275W}-m_{\rm F438W}$ CMD of the surviving MSa stars that passed
both selections of panels (a) and (b). We removed a few additional
outliers from our MSa candidates (black points outside the two green
lines).  Note that on this panel we can clearly distinguish a
secondary, less-populated sequence of stars to the red side of the
main distribution. As we have shown in \citet[their Fig.~13 and
  related discussion]{2010AJ....140..631B}, these stars are not
binaries and constitute a distinct subpopulation of the MSa. In fact,
for $m_{\rm F438W}<20$, The MSa sequence is almost vertical, and any
binary sequence made up of MSa stars is going to necessarily merge
with the MSa stars themselves on a CMD. What we see in panel (c) (as
well as in panel a), instead, is that the secondary population runs
parallel to MSa stars also for $m_{\rm F438W}<20$.

``Chromosome'' maps are a very powerful tool that have been used
extensively over the last few years to reveal multiple-population
substructures in GCs (see, e.g., \citealt{2015MNRAS.447..927M,
  2015ApJ...808...51M, 2017MNRAS.464.3636M}, and references therein).
Briefly, the construction of a chromosome map begins by tracing two
guide lines enclosing a given population in a CMD based on a
particular color (e.g., $m_{\rm F336W}-m_{\rm F438W}$). These two
lines are then used to rectify and parallelize the population
sequence\footnote[1]{Basically, to the color of each star we subtract
  the color of the blue-boundary line at the same magnitude level of
  the star, and then we divide this quantity by the difference in
  color between the red- and the blue-boundary lines (again at the
  same magnitude level).}, which then appears as a vertical
distribution of stars of constant width in the $\Delta^{\rm N}_{\rm
  F336W-F438W}$ pseudo CMD (where $\Delta^{\rm N}_{\rm F336W-F438W}$
indicates the rectified and parallelized pseudo color). In the pseudo
CMD, the bluer and redder guide lines used in the rectification and
parallelization process (in other words, a homographic transformation
of the CMD plane) are transformed into vertical lines at abscissa 0
and 1, respectively (see \citealt{2015MNRAS.447..927M,
  2015ApJ...808...51M} for more details). The same procedure can be
applied to a CMD based on a different color, e.g. $m_{\rm
  F275W}-m_{\rm F336W}$ to obtain the pseudo-color $\Delta^{\rm
  N}_{\rm F275W-F336W}$. The two pseudo colors derived this way are
then plotted one against the other to create a chromosome map.  (Note
that chromosome maps can also be constructed using color indexes
instead of colors. A color index is defined as a difference between
two colors with a filter in common, e.g.:\ $C_{\rm
  F275W,F336W,F438W}$=$(m_{\rm F275W}-m_{\rm F336W}) - (m_{\rm
  F336W}-m_{\rm F438W})$, where the filter in common in this case is
F336W, see also \citet{2013ApJ...767..120M,2015AJ....149...91P} and
references therein.)

The $\Delta^{\rm N}_{\rm F275W-F336W}$ vs.\ $\Delta^{\rm N}_{\rm
  F336W-F438W}$ chromosome map of selected MSa stars is shown in panel
(d) of Fig.~\ref{f:msa} (black points). In both the $m_{\rm
  F275W}-m_{\rm F336W}$ and $m_{\rm F336W}-m_{\rm F438W}$ CMDs we
simulated a single sequence of stars, the color spread of which is
defined by photometric and differential-reddening-correction errors
only, and passing in between the two lines that were used to rectify
and parallelize the population sequence in each CMD.  We computed the
$\Delta^{\rm N}_{\rm F275W-F336W}$ vs.\ $\Delta^{\rm N}_{\rm
  F336W-F438W}$ values for this simulated sequence to show how the
chromosome map of a single sequence would appear (red points in panel
d). The black ellipse encloses the 68.27 percentile of the simulated
stars. It is clear that the chromosome map of selected MSa stars is
much wider (and clumpy) than what photometric and
differential-reddening-correction errors would predict.

The Hess diagram of the chromosome map of selected MSa stars is shown
in panel (e). The color mapping of this and the following Hess
diagrams goes from purple (lowest density) to blue, green (average
density), yellow and red (highest density). The Hess diagram is
provided with the only purpose of giving the reader a qualitative and
clearer sense of the distribution of stars in the chromosome map.  In
panels (d) and (e) we can clearly distinguish two clumps:\ a main
clump at about (0.45, 0.45), with a tail extending down to (0.1, 0.0),
and a second, less-populated clump located at about (0.9,0.9).  The
tail of the main clump could be a hint of a spread in light-element
abundances of the main-clump subpopulation, or it could even represent
a distinct subpopulation of stars partially overlapping the main clump
in the chromosome map. On the other hand, the tail could also be the
result of a non-optimal subtraction of outliers. Given the uncertain
nature of the tail, in panel (f) we conservatively defined only two
subpopulations of MSa:\ the MSa1 (dark yellow) and the MSa2 (light
yellow), within the black envelopes. Stars outside these black
envelopes are rejected. The selections defined in panel (f) set the
final sample of the MSa subpopulations.  This analysis confirms the
findings of \citet{2010AJ....140..631B} that the MSa population is
split into two components.  The inclusion of the tail into the MSa1
selection will have little or no effect on the average properties of
the MSa1 subpopulation, given the relatively small number of stars in
the tail with respect to the total number of MSa1 stars.

\begin{figure*}[!t]
\centering
\includegraphics[angle=-90,width=16cm]{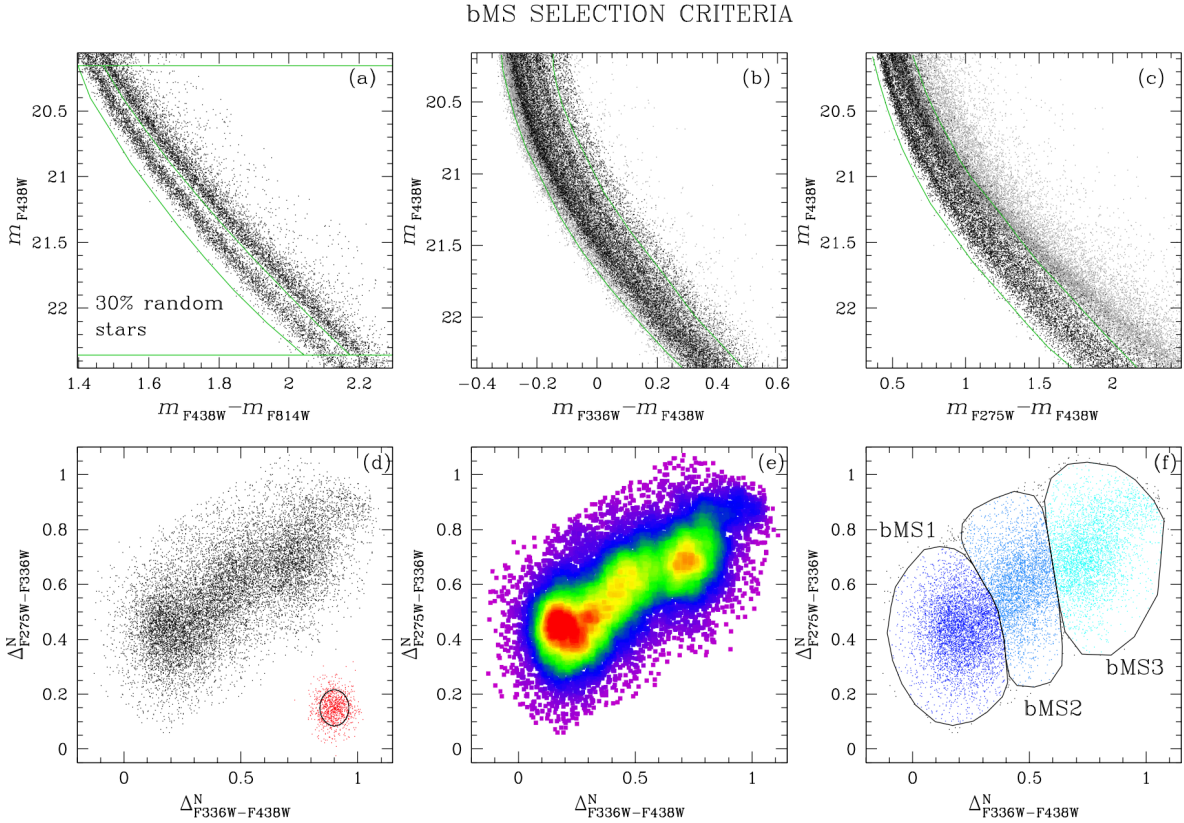}
\vskip -.3cm
\caption{\small{These six panels are arranged in a similar way as
    those of Fig.~\ref{f:msa}, but this time we show how we selected
    bMS stars. We preliminary selected bMS stars on the $m_{\rm
      F438W}-m_{\rm F814W}$ CMD in panel (a), within the green
    lines. This particular CMD offers the clearest separation between
    bMS and rMS stars. For clarity, we plotted a randomly-selected
    30\% of the stars. Already identified MSa1 and MSa2 stars have
    been removed from the CMD. Preliminary-selected bMS stars are
    further refined using the $m_{\rm 336W}-m_{\rm F438W}$ and $m_{\rm
      F275W}-m_{rm F438W}$ shown in panels (b) and (c). As we did for
    Fig.~\ref{f:msa}, survived stars from the previous panel are in
    black, while rejected stars are in gray. $\Delta_{m_{\rm
        F275W}-m_{\rm F336W}}^{\rm N}$ vs.\ $\Delta_{m_{\rm
        F336W}-m_{\rm F438W}}^{\rm N}$ chromosome map and Hess diagram
    are in panels (d) and (e), respectively. Panel (e) reveals at
    least three main subcomponents of the bMS, that we identify as
    bMS1 (dark blue), bMS2 (azure), and bMS3 (light blue) in panel
    (f).\\~\\}}
\label{f:bms}
\end{figure*}

\begin{figure*}[!t]
\centering
\includegraphics[angle=-90,width=16cm]{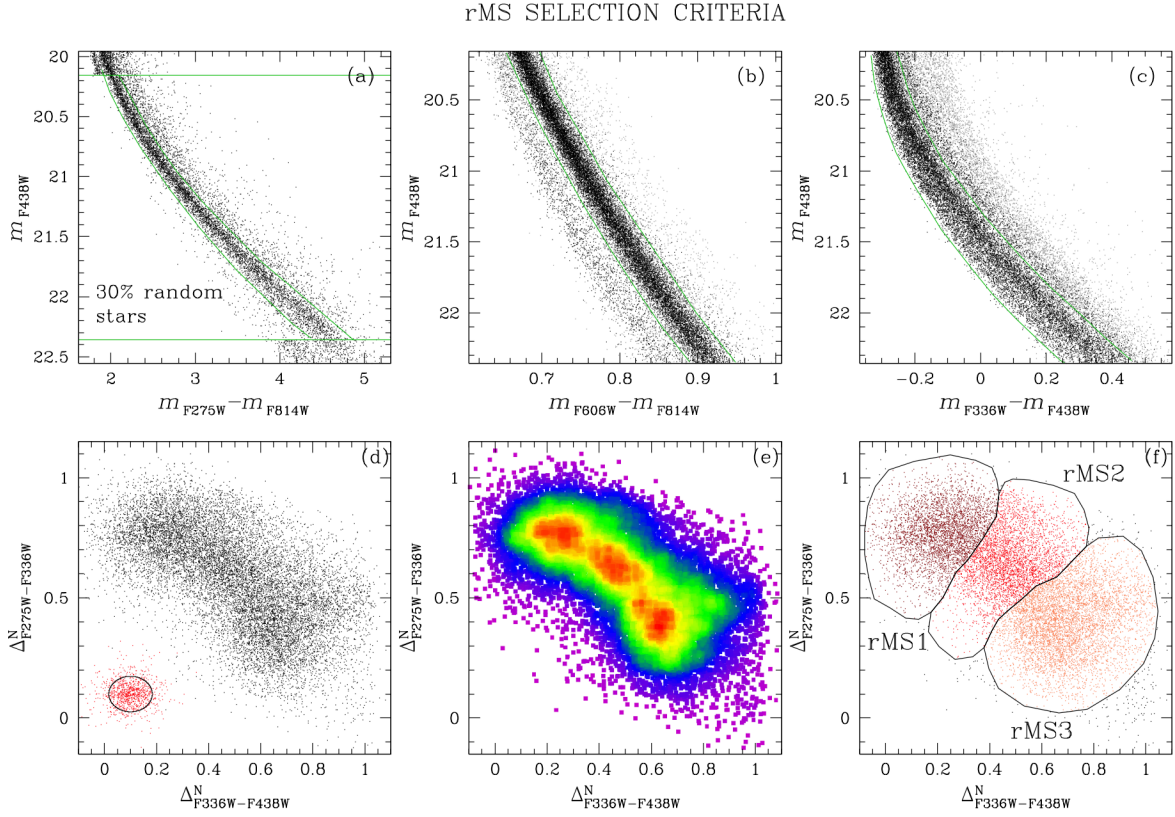}
\vskip -.3cm
\caption{\small{Similar to Fig.~\ref{f:msa}, but for the rMS. As for
    the bMS, the Hess diagram in panel (e) reveals at least three main
    subcomponents, labeled as rMS1 (brown), rMS2 (red), and rMS3
    (orange) in panel (f).\\~\\}}
\label{f:rms}
\end{figure*}

\begin{figure*}[!t]
\centering \includegraphics[angle=-90,width=15cm]{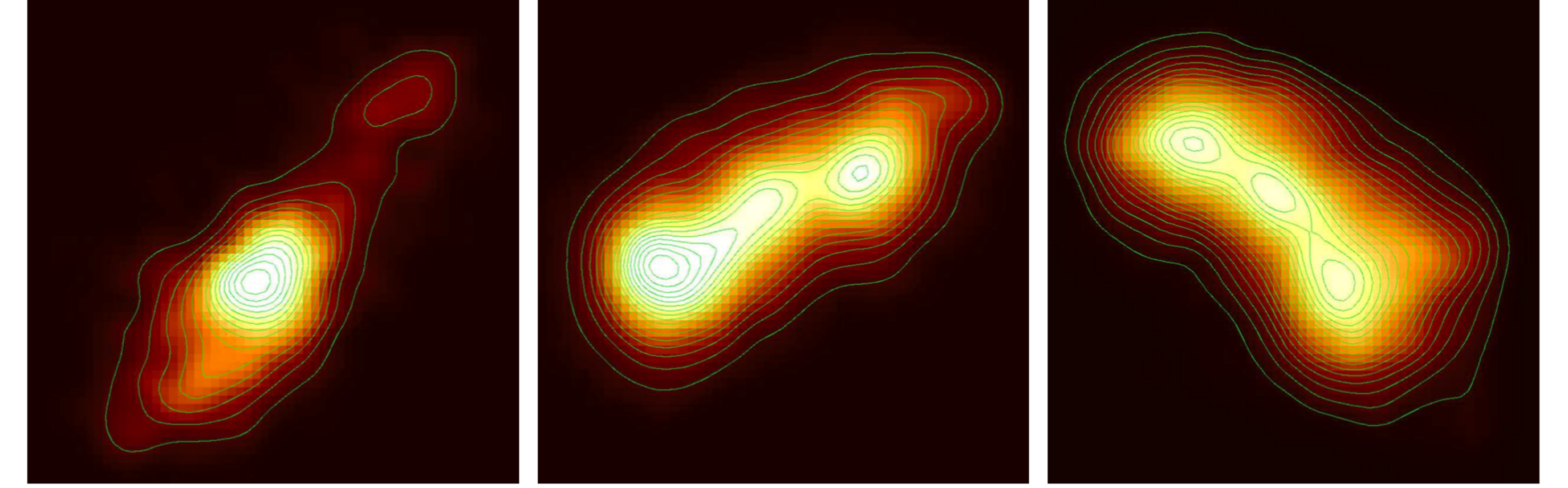}
\vskip 2mm
\caption{\small{Gaussian-smoothed $\Delta_{m_{\rm F275W}-m_{\rm
        F336W}}^{\rm N}$ vs.\ $\Delta_{m_{\rm F336W}-m_{\rm
        F438W}}^{\rm N}$ chromosome maps for the MSa (left), the bMS
    (middle), and the rMS (right) subpopulations. Isodensity contours
    are shown in green.  Axes quantities, scales and ranges are the
    same as the corresponding panels (d), (e) and (f) of each
    population selection.  \\~\\}}
\label{f:recap}
\end{figure*}

\subsection{The bMS}
\label{ss:bms}

The second easiest population to isolate is the bMS, which typically
stands out quite clearly on the blue side of the MS in most CMDs of
\wcen, in particular in the $m_{\rm F438W}-m_{\rm F814W}$ CMD shown in
panel (a) of Fig.~\ref{f:bms}. On this and the following panels of the
figure, all previously-identified stars (in this case, MSa1 and MSa2
stars) have been removed, in order to facilitate the selection of the
remaining populations. For clarity, in panel (a) we plotted a
randomly-selected 30\% of the stars.

To select bMS stars, we kept the same faint limit as for the MSa, i.e.
$m_{\rm F438W}=22.36$, but we had to lower the bright limit to $m_{\rm
  F438W}=20.16$.  In fact, at magnitudes brighter than $m_{\rm
  F438W}=20.16$, the bMS and the rMS become too close to each other
(and eventually overlap below the turn off) in every CMD regardless of
the adopted color baseline.  As we did for the MSa, we drew by hand
two lines (also in green) defining the color boundaries of our
preliminary-selected bMS stars. At this stage, both boundaries are
purposely generous and necessarily enclose rMS contaminants. We want
to be as inclusive as possible in our first selection, so to start
from a sample containing most --if not all-- bMS stars. Contaminants
will be rejected later using CMDs based on different colors, as we
have done for MSa stars.

In panel (b) of Fig.~\ref{f:bms} we show the $m_{\rm 336W}-m_{\rm
  F438W}$ CMD of preliminary-selected bMS stars from panel (a). Stars
rejected in panel (a) are in gray.  The bMS on this CMD is
significantly spread in color (with hints of substructures). A small
fraction of stars are smeared toward redder colors. We removed this
red tail of stars, together with a few other outliers on the blue
side, using the two green lines.  A further selection refinement is
applied on the $m_{\rm F275W}-m_{rm F438W}$ CMD (panel c) .  Note that
the bulk distribution of bMS stars is also spread in panel (c), with
hints of a split.

The $\Delta^{\rm N}_{\rm F275W-F336W}$ vs.\ $\Delta^{\rm N}_{\rm
  F336W-F438W}$ chromosome map of survived bMS stars after the
selections we applied in panels (a)+(b)+(c) is shown in black in panel
(d). In red the expected distribution of a single sequence of stars,
based on only photometric and differential-reddening-correction
errors. The ellipse encloses the 68.27 percentile of the
single-sequence distribution. The Hess diagram of the chromosome map
is presented in panel (e). Three main clumps of stars can be clearly
seen in these two panels.  All of them exhibit some degree of
substructures. In particular, the bluer clump, located at (0.2, 0.4),
shows a red tail that pushes out towards the central clump, while
another tail of points emerges from the redder clump, located at
(0.75,0.7) and runs out to about (1.1,0.9).

Following the same conservative approach we applied to MSa stars, we
defined three subpopulations of the bMS in panel (f):\ the bMS1
(blue), the bMS2 (azure), and the bMS3 (cyan), each defined as all
stars within the respective black envelope.

The identification of the bMS2 might appear less obvious than that of
the other two bMS subpopulations, and one could argue that the bMS2 is
simply part of the tail of the distribution of bMS1 stars. It should
be noted, however, that the overdensity at the center of the bMS2
distribution is about 3 sigmas away from the peaks of both the bMS1
and the bMS3, in terms of photometric and differential-reddening
errors alone. Moreover, if errors alone are the cause of the central
overdensity that we identify as the bMS2 --- which is redder than the
bMS1 in the $m_{\rm F336W}-m_{rm F438W}$ CMD---, then stars of the
``bMS2'' would have an equal chance of being bluer or redder than the
bMS1 in another CMD based on different filters. We see, instead, that
bMS2 stars are systematically bluer that bMS1 stars in the $m_{\rm
  F275W}-m_{\rm F336W}$:\ a particular characteristic that implies a
different light-element abundance for the bMS1 and the bMS2.  Finally,
it could be that the bMS1 population formed over a relatively long
period of time, and its stars contain a spread ---as opposed to a
split in the case of two short star-formation bursts--- in
light-element abundance:\ spread that we have identified as the bMS2.

All of these caveats notwithstanding, for the time being let us
consider the bMS1 and the bMS2 as two distinct subpopulations.

It stands to reason that bMS1 and bMS3 stars will necessarily be
contaminated by some bMS2 stars, and vice versa. Nevertheless, the
average properties of stars within each subpopulation selection will
still be representative of the subpopulation itself.

\subsection{The rMS}
\label{ss:rms}

Once bMS and MSa stars have been removed, rMS stars stand out quite
clearly in the CMD, e.g. in the $m_{\rm F275W}-m_{\rm F814W}$ CMD we
show in panel (a) of Fig.~\ref{f:rms}. For clarity, in this panel we
plotted a randomly-selected 30\% of the stars, as we did in panel (a)
of Fig.~\ref{f:bms}.  We kept the same magnitude limits as for the bMS
(green horizontal lines), and preliminary selected (by hand) rMS stars
on this panel by means of the two diagonal green curves.

Selected stars are then plotted in black in the $m_{\rm F606W}-m_{\rm
  F814W}$ CMD of panel (b).  What appeared as a single sequence in
panel (a) is now clearly split into two sequences: a well defined,
more-populated sequence on the red side and a less-populated sequence
on the blue side of the bulk population. The $m_{\rm F606W}-m_{\rm
  F814W}$ CMD, which is mostly unaffected by light-element-abundance
variations, helps us in distinguishing populations with different He
enhancement. The sequence of stars on the blue side of the bulk
population is therefore likely to be He-enhanced with respect to the
bulk population itself.\footnote[2]{Note that MSa and rMS stars are
  overlapped in the $m_{\rm F606W}-m_{\rm F814W}$ CMD, with MSa stars
  being mostly parallel to bMS stars. More in Sect.~\ref{s:ov}.} Since
the rMS should be made up of first-generation, Fe-poor and He-normal
stars (\citealt{2005ApJ...621..777P}) we removed stars on this bluer
sequence from our rMS selection, together with a few outliers on the
red side of the bulk population. We will return to this blue sequence
in the next subsection.  An additional rejection of likely outliers is
performed on the $m_{\rm F336W}-m_{\rm F438W}$ CMD of panel (c).

Panel (d) shows the $\Delta_{m_{\rm F275W}-m_{\rm F336W}}^{\rm N}$
vs.\ $\Delta_{m_{\rm F336W}-m_{\rm F438W}}^{\rm N}$ chromosome map of
selected rMS stars (in black) and the error distribution (in red). The
corresponding Hess diagram is in panel (e). As for the bMS, three main
clumps of stars stand out clearly in the chromosome map, with the
rightmost clump, located at (0.6, 0.3) possibly showing a
scarcely-populated tail of stars extending towards (0.9,0.0).  Again,
we conservatively defined just three rMS subpopulations in panel
(f):\ rMS1 (brown), rMS2 (red), and rMS3 (orange).  As for the bMS
subpopulations, any slight cross-contamination of the rMS
subpopulations should not affect their general properties.

It is worth noting that ---in contrast to the behavior of MSa and bMS
stars in the $\Delta_{m_{\rm F275W}-m_{\rm F336W}}^{\rm N}$
vs.\ $\Delta_{m_{\rm F336W}-m_{\rm F438W}}^{\rm N}$ chromosome map,
which both showed populations aligned from the bottom-left to the
upper-right, here we see that the rMS stars are aligned from the
bottom-right to the upper-left. We will return to this property in
Sect.~\ref{s:ana}.

\subsection{Hidden MS populations}
\label{ss:msx}

In the previous subsections, we identified stars belonging to the
three main populations of \wcen, namely:\ the MSa, the bMS, and the
rMS. Between $20.16\le m_{\rm F438W}\le 22.36$ ---the magnitude
interval we used to select both the bMS and the rMS--- we have a total
39$\,$529 high-photometric-quality MS stars.  In this magnitude range,
the selected MS populations (and their subpopulations) account for:\
\begin{itemize}
\item{MSa:\ 1394 stars ($3.53\pm 0.10$\%), of which 1283 stars
  ($3.25\pm 0.09$\%) belong to the MSa1 subpopulation, and 111 stars
  ($0.28\pm 0.03$\%) are MSa2 stars.}
\item{bMS:\ 12\,776 stars ($32.32\pm 0.33$\%), so divided:\ 5141
  ($13.01\pm 0.19$\%) bMS1 stars, 3683 ($9.32\pm 0.16$\%) bMS2 stars,
  and 3952 ($10.00\pm 0.17$\%) bMS3 stars.}
\item{rMS:\ 13\,124 stars ($33.20\pm 0.33$\%), divided into 3739
  ($9.46\pm 0.16$\%) rMS1 stars, 3838 ($9.71\pm 0.16$\%) rMS2 stars,
  and 5547 ($14.03\pm 0.20$\%) rMS3 stars.}
\end{itemize}
The quoted errors correspond to Poisson errors only.
Figure~\ref{f:recap} provides an alternative view of the chromosome
maps of the three populations. Panels from left to right show a
bichromatic (red to yellow, low to high) Gaussian-smoothed version of
panels (e) of Figs.~\ref{f:msa}, \ref{f:bms} and \ref{f:rms},
respectively. Isodensity contours are also shown, for clarity. These
Gaussian-smoothed plots emphasize better some subtle features present
in the chromosome maps, and should be used together with the
chromosome maps and the Hess diagrams in panels (d) and (e) of
Figs.~\ref{f:msa}, \ref{f:bms} and \ref{f:rms} to qualitatively assess
the efficacy of our TRSR method.

Overall, selected populations account for only $\sim$69\% of MS stars.
There are still 12\,229 unidentified stars (a number comparable in
size to that of bMS or rMS stars!)  that seemingly do not belong to
either the MSa, the bMS, or the rMS. Our TRSR method is limited to
only four passages, hence selection effects alone are unlikely to be
the cause of such a large number of still unidentified stars.  When we
selected rMS stars in panel (b) of Fig.~\ref{f:rms}, it was clear that
we rejected stars belonging to a previously-unidentified population.
Possibly, something similar --but less obvious-- also happened when we
rejected what we thought were bMS outliers in panel (b) of
Fig.~\ref{f:bms}.

So far, we have identified a total of 8 subpopulations in \wcen, which
already make this cluster the most complex of all GCs. Nevertheless,
it seems that more is left to unravel. Our TRSR method proved to work
reasonably well so far. Now, what happens if we remove all 8
subpopulations from a CMD and apply the same selection procedures to
what is left?

\begin{figure*}[!t]
\centering
\includegraphics[width=16cm]{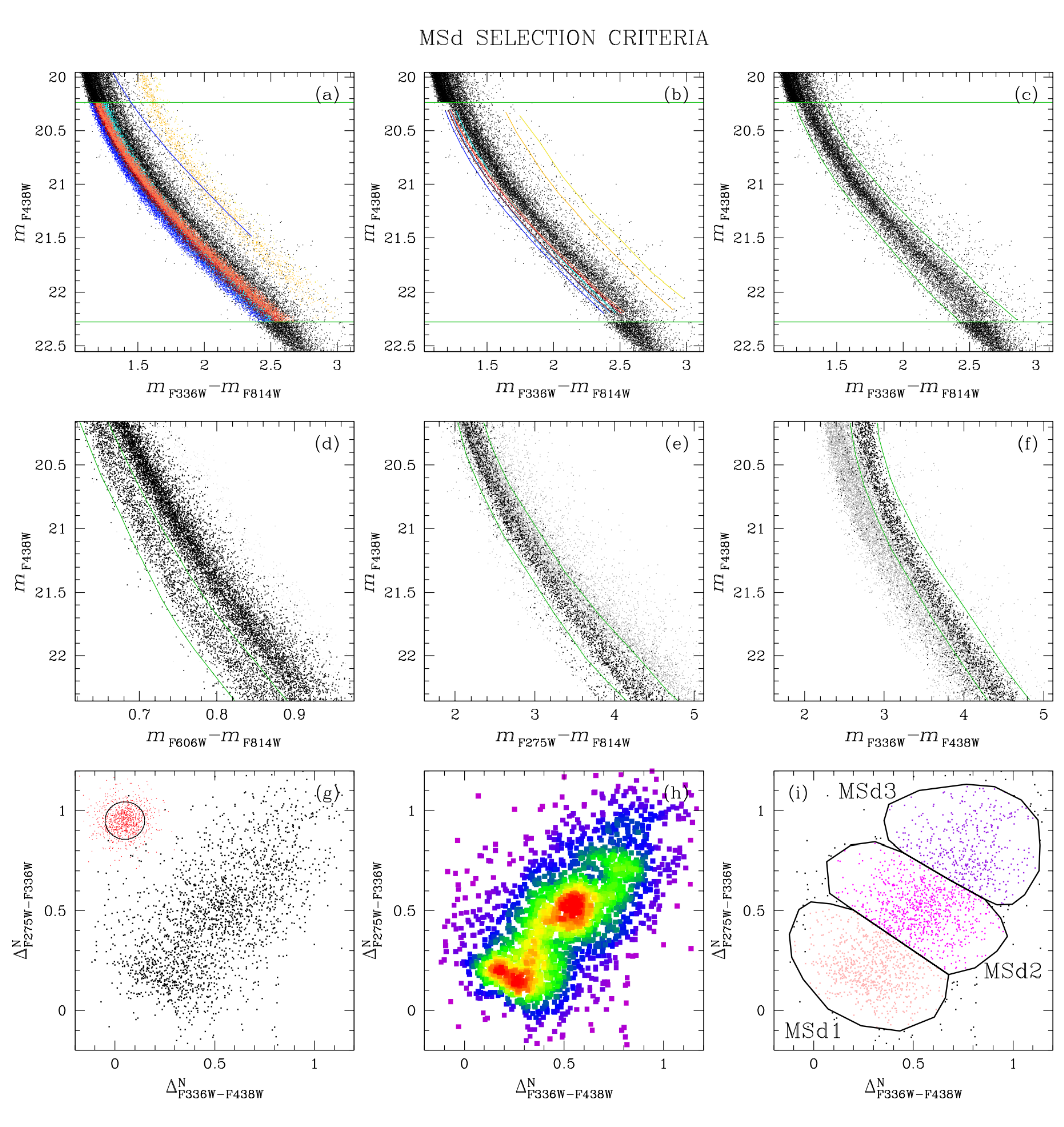}
\caption{\small{(a) $m_{\rm F336W}-m_{\rm F814W}$ CMD of the MS of
    \wcen.  Stars belonging to previously-identified subpopulations
    are color-coded accordingly.  Unidentified stars are in
    black. Panel (b) shows the same CMD as in panel (a), in which now
    color-coded fiducial lines replace identified stars.  Panel (c) is
    similar to that of panels (a) and (b), but now only unidentified
    stars are shown. On this panel, we preliminary selected stars that
    seem to form a well-defined sequence. These stars are plotted in
    (d) on the $m_{\rm F606W}-m_{\rm F814W}$ CMD, where they clearly
    split into two components. On the rest of the figure we focus on
    the less-populated blue component, that we hereafter call as the
    MSd. Panels (e) and (f) illustrate our selection refinements for
    MSd stars.  The chromosome map (g) and Hess diagram (h) of MSd
    stars reveal three main subpopulations, that we define in panel
    (i) as MSd1 (pink), and MSd2 (magenta), and MSd3 (purple).\\~\\}}
\label{f:msd}
\end{figure*}

\begin{figure*}[!t]
\centering
\includegraphics[angle=-90,width=16cm]{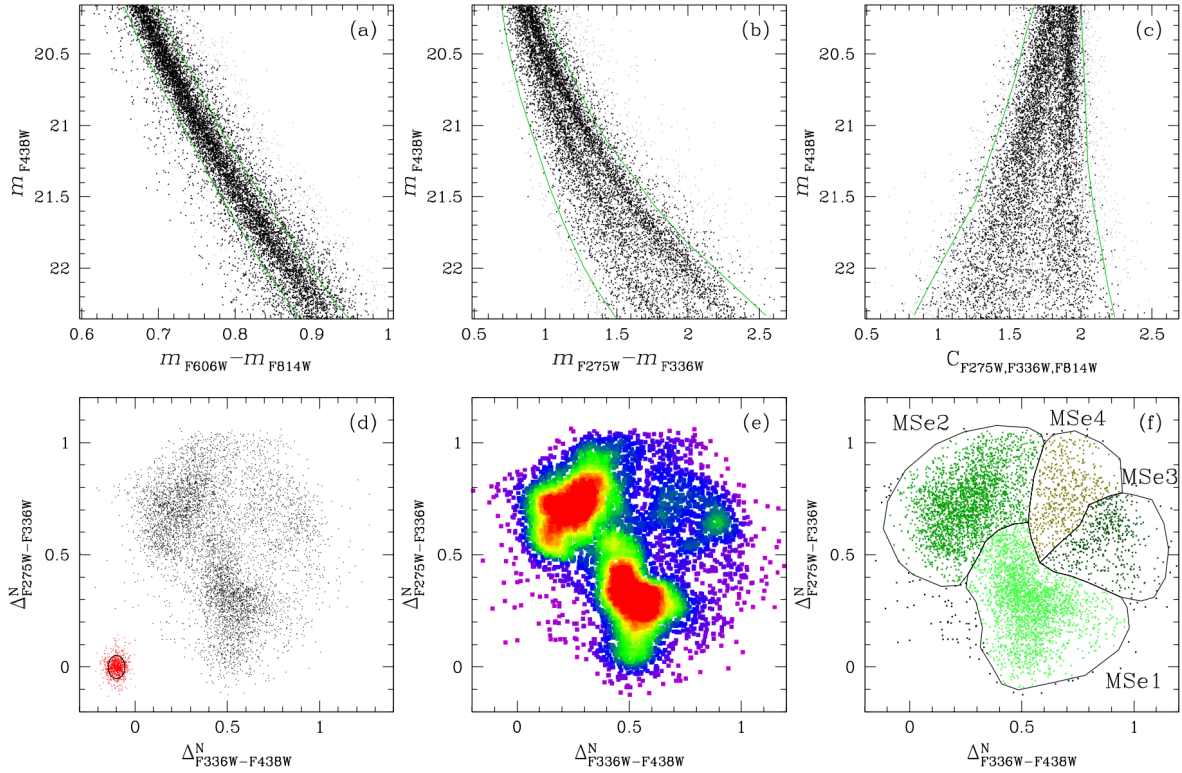}
\vskip -3mm
\caption{\small{Panel (a) is a replica of panel (d) of
    Fig.~\ref{f:msd}, in which we also removed MSd stars. The
    remaining stars (hereafter MSe) form a well-defined sequence on
    this plane, which we select and further refine in panels (b) and
    (c). Note that in both panels (b) and (c) MSe stars appear to be
    split into two sequences. Also note that in panel (c) we use the
    C$_{\rm F275W,F336W,F814W}$ pseudo color CMD instead of a normal
    CMD. The chromosome map and the Hess diagram shown in panels (d)
    and (e), respectively, reveal a quite complex picture, with two
    main clumps of stars with asymmetric shape and two additional
    less-populated clumps that occupy a well-defined region. These
    four clumps are identified in panel (f) as:\ MSe1 (lime), MSe2
    (green), MSe3 (dark green) and MSe4 (olive).\\~\\}}
\label{f:mse}
\end{figure*}

\begin{figure*}[!t]
\centering
\includegraphics[angle=-90,width=12cm]{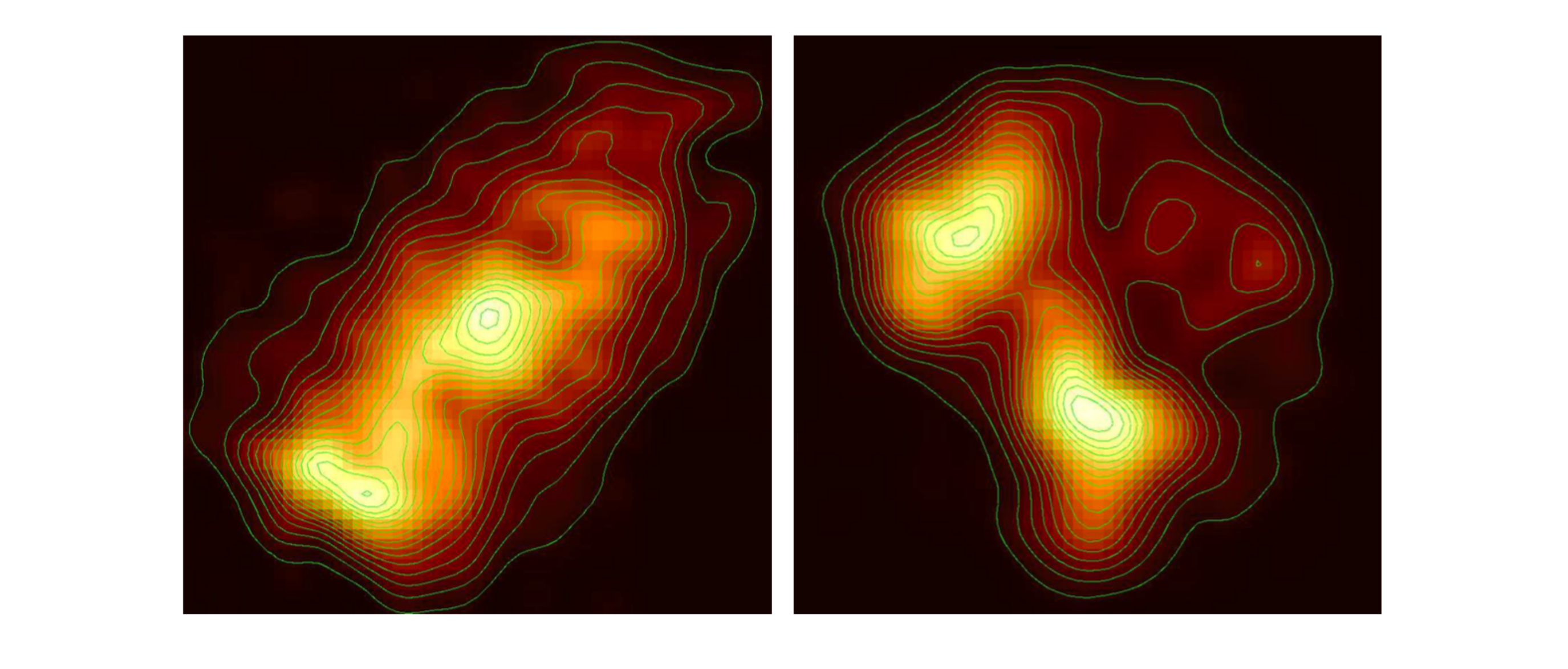}
\caption{\small{Gaussian-smoothed $\Delta_{m_{\rm F275W}-m_{\rm
        F336W}}^{\rm N}$ vs.\ $\Delta_{m_{\rm F336W}-m_{\rm
        F438W}}^{\rm N}$ chromosome maps for the MSd (left), the MSe
    (right) subpopulations. Isodensity contours are shown in
    green. Axes quantities, scales and ranges are the same as the
    respective panels (d), (e) and (f) of each population
    selection.\\~\\}}
\label{f:recap2}
\end{figure*}

\subsection{The MSd}
\label{ss:msd}

Besides MSa1 and MSa2 stars, the remaining 6 subpopulations (bMS1,
bMS2, bMS3, rMS1, rMS2 and rMS3) are mostly overlapped with each other
in the $m_{\rm F336W}-m_{\rm F814W}$ CMD. We show this CMD in panel
(a) of Fig.~\ref{f:msd}. Stars belonging to previously-identified
subpopulations are color-coded accordingly. The two horizontal green
lines mark the magnitude interval used to select bMS and rMS
stars. The vast majority of unidentified stars on this CMD lie to the
red side of both bMS and rMS subpopulations, and to the blue side of
MSa stars.

In order to show that these unidentified stars are unlikely to be all
unresolved binaries and/or blends, we marked with a blue line the
locus of equal-mass bMS1-bMS1 binaries, being bMS1 stars the
bluest/faintest among the 8 identified subpopulations. Equal-mass
binaries made up by any other subpopulation combination are expected
to be brighter/redder than this blue line.  Typically, the mass-ratio
distribution of binary stars is found to be almost flat for most of
globular clusters, with an overdensity near the equal-mass binary
sequence (\citealt{2012A&A...540A..16M}). This translates into a
close-to-uniform distribution of binaries between a single-star
sequence and the equal-mass binary sequence.

What we see in panel (a) of Fig.~\ref{f:msd} is, instead, that
still-unidentified stars lie in close proximity of the bulk of the MS,
and significantly far from the blue line. Unless the vast majority of
binaries in \wcen\ are made up of stars where the primary is always
far more massive than the secondary ---an extremely unlikely event
indeed---, it is apparent that the unidentified stars cannot be all
binaries.

The blend hypothesis can also be quickly dismissed. Our star list
contains high-quality photometric measurements in five filters, from
the near UV to the \textit{I} band. Let us suppose that two stars are
so close to each other on the FoV that our reduction software measured
only one position and magnitude for them. Then, unless these two stars
have a very similar luminosity in all five filters (again, a very
unlikely event), the measured position of the blend will be much
closer to the hotter source in F275W, and much closer to the colder
source in F814W, because of the negligible contribution of the other
source in these bands. This would result in an artificially increased
positional rms for the blend, and the blend would have been rejected
in our PM selections. Even in the case where the two sources have a
similar luminosity in one band, then the \texttt{QFIT} parameter would
tell us that the PSF fit was not optimal, and the blend would have
been rejected in our photometric selections. Of course, we cannot
exclude that a few blends survived all the astro-photometric
selections we applied and made it into our final star list, but these
blends are expected to be a rarity, not 30\% of the MS stars.

Panel (b) of Fig.~\ref{f:msd} is similar to panel (a), except that
stars of the 8 previously-identified subpopulations have been removed
and replaced by the subpopulation fiducial lines (color-coded
accordingly).  Fiducial lines are obtained via least-squares fitting
of a third-order polynomial to the stars of each population on a CMD.
In panel (c) we show the same CMD of panels (a) and (b), but for
unidentified stars only. These stars appear to form a single,
well-defined sequence on the $m_{\rm F336W}-m_{\rm F814W}$ CMD, which
we preliminarily select by means of the two green diagonal lines
(drawn by hand).

Panel (d) shows the $m_{\rm F606W}-m_{\rm F814W}$ CMD of these
selected stars in black. Surprisingly, what seemed to be a single
sequence in panel (c) now splits into two components. A quick
cross-check revealed that the blue component is made up of the same
stars that we rejected during our rMS selection procedures (panel b of
Fig.~\ref{f:rms}). Despite the similarities between panel (d) of
Fig.~\ref{f:msd} and panel (b) of Fig.~\ref{f:rms}, the red component
in Fig.~\ref{f:msd} is \textit{not} made by rMS stars, as rMS stars
have already been removed from the CMD. The red component we see in
panel (d) of Fig.~\ref{f:msd} and rMS stars simply happen to overlap
each other in the $m_{\rm F606W}-m_{\rm F814W}$-baseline CMD.

Since the $m_{\rm F606W}-m_{\rm F814W}$ color is a clear tracer of He
and Fe abundance variations, the two components we see in panel (d)
are very likely to have different He abundances and/or different
metallicity. They exhibit a similar behavior to that of bMS and rMS
stars, i.e., they are clearly split in $m_{\rm F606W}-m_{\rm F814W}$,
but are overlapped in $m_{\rm F336W}-m_{\rm F814W}$.

First, we focused on the blue component, which we will hereafter call
as ``MSd''. MSd stars are preliminarily selected (green lines) in
panel (d). We further refined the MSd sample by removing a few
outliers using the $m_{\rm F275W}-m_{\rm F814W}$ and $m_{\rm
  F336W}-m_{\rm F438W}$ CMDs (panels (e) and (f), respectively).

The chromosome map and the Hess diagram of selected MSd stars are
shown in panels (g) and (h), respectively. The Hess diagram highlights
two main clumps of stars located at (0.25,0.15) and (0.55,0.5), and a
less-populated clump at (0.8,0.7). We identified these three clumps in
panel (i) as the subpopulations MSd1 (pink), MSd2 (magenta) and MSd3
(purple).  Another hint about the possible chemical similarity between
MSd and bMS stars is given by the similar orientation of the
respective subpopulations on the chromosome map.

The identification of the MSd3 subpopulation is somewhat less obvious
than that of the other two MSd subpopulations, and the same arguments
that we made for the bMS2 can be applied to the MSd3 as well.

\subsection{The MSe}
\label{ss:mse}

Now, let us focus our attention on the red component (hereafter MSe)
we left aside in panel (d) of Fig.~\ref{f:msd}. Panel (a) of
Fig.~\ref{f:mse} is similar to panel (d) of Fig.~\ref{f:msd}, but here
we have removed MSd stars. In this panel we preliminary selected MSe
stars as those within the two green lines, and we further refined the
MSe sample as shown in panels (b) and (c). Note that this time panel
(c) shows the pseudo-CMD based on the color index $C_{\rm
  F275W,F336W,F814W}$, instead of a normal CMD as it was the case for
the previous figures.  In both panels (b) and (c), MSe stars clearly
split into two predominant sequences.

Panels (d) and (e) of Fig.~\ref{f:mse} show the chromosome map of
selected MSe stars and its Hess diagram, in which the two sequences we
saw in panels (b) and (c) stand out clearly as the two main clumps,
located at about (0.2,0.7) and (0.5,0.3). It is worth noting that the
shape of both these clumps is far from being symmetric:\ a possible
indication of substructures. There are also two additional, less
populated clumps centered at around (0.65,0.7) and (0.9,0.65). Since
these two lesser clumps are located in distinct regions of the
chromosome map, significantly far from the two main components, we
propose that they actually constitute two additional subpopulations of
the MSe (this can be better seen in Fig.~\ref{f:recap2}). In panel (f)
we therefore selected the following four MSe subpopulations:\ MSe1
(lime), MSe2 (green), MSe3 (dark green), and MSe4 (olive).

Please note that the single-population error distribution (red points
in panel d of Fig~\ref{f:mse}) is comparable in size to that of the
MSe3 and the MSe4, but significantly smaller than the MSe1 and the
MSe2, further suggesting that what we have defined as MSe1 and MSe2
could likely hide additional subcomponents. Moreover, both the Hess
diagram of the chromosome map (panel e of Fig~\ref{f:mse}) and its
Gaussian-smoothed representation (right panel of Fig.~\ref{f:recap2}
do not do justice to highlight the actual significance of the MSe3 and
MSe4 peaks. We employed a linear color-mapping for the Hess diagrams
and a linear isodensity contour spacing for the Gaussian-smoothed maps
of all the previous populations. For consistency, we also adopted the
same scheme for the MSe. Since the MSe subpopulation relative fraction
can vary by up to a factor of 5, a logarithmic color mapping and
contour spacing might have been more appropriate. Finally, we remark
that MSe3 and MSe4 stars have, on average, the same photometric errors
as MSa2 stars, which are again about a factor of five less numerous
that both the MSe3 and the MSe4.

In summary, we have discovered 2 major new populations on the MS of
\wcen, which were previously hidden by bMS and rMS stars. We call
these new populations MSd and MSe. The MSd is in turn made up of three
subpopulations (MSd1, MSd2 and MSd3), and shares similar properties to
bMS stars. The MSe is made up of four subpopulations and shares
similar properties to rMS stars. The MSd and MSe populations account
for:\
\begin{itemize}
\item{MSd:\ 2016 stars ($5.10\pm 0.12$\%), of which 757 ($1.92\pm
  0.07$\%) are MSd1 stars, 819 ($2.07\pm 0.07$\%) are MSd2 stars, and
  440 ($1.11\pm 0.05$\%) are MSd3 stars.}
\item{MSe:\ 6129 stars ($15.51\pm 0.21$\%), subdivided into 2555
  ($6.46\pm 0.13$\%) MSe1 stars, 2591 ($6.56\pm 0.13$\%) MSe2 stars,
  463 ($1.17\pm 0.05$\%) MSe3 stars, and 520 ($1.32\pm 0.06$\%) MSe4
  stars.}
\end{itemize}

As we have done for the three canonical MS populations of \wcen\ (MSa,
rMS and bMS), we show in Fig.~\ref{f:recap2} a Gaussian-smoothed
version of the $\Delta_{m_{\rm F275W}-m_{\rm F336W}}^{\rm N}$
vs.\ $\Delta_{m_{\rm F336W}-m_{\rm F438W}}^{\rm N}$ chromosome maps
for the MSd (left), the MSe (right) subpopulations. Isodensity
contours are shown in green.

\begin{figure*}[!t]
\centering
\includegraphics[angle=-90,width=16cm]{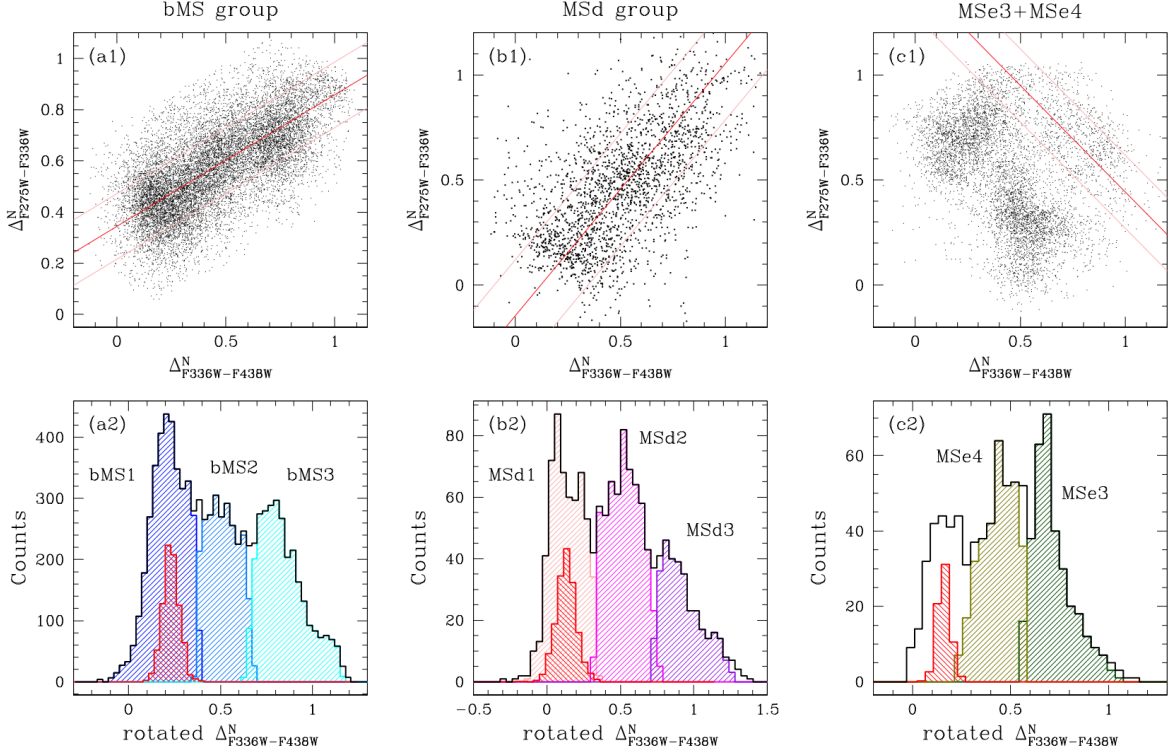}
\vskip -1cm
\caption{Top panels:\ replicas of the chromosome map for the bMS
  (left), the MSd (middle), and the MSe (right). In each panel, we
  fitted a straight line (in red) to the barycenters of the
  subpopulations within (only to the MSe3 and the MSe4 for the MSe),
  and rotated the plane in such a way that the fitted line becomes
  parallel to the abscissa. The rotation point is arbitrary:\ we chose
  the center point at (0.5, 0.5) in the chromosome map. We constructed
  a histogram (in black, bottom panels) using all stars within
  $\pm$1$\sigma$ along the ordinate distribution (pink lines). The
  histogram of each subpopulation, extracted within the pink lines, is
  color-coded. The histogram of photometric plus
  differential-reddening-correction errors only is shown in red. See
  the text for details.  \\~\\}
\label{f:histo}
\end{figure*}

\subsection{A brief discussion on the less-constrained subpopulations}
\label{ss:secure}

In the previous subsections, we selected four subpopulations,
namely:\ bMS2, MSd3, MSe3 and MSe4, for which the identification was
less obvious than for the other subpopulations. In order to give the
reader an additional point of view on this subject, we extracted
histograms of the subpopulations of the bMS, MSd and MSe on the
chromosome map as follow. Let us take the bMS as an example. We fitted
a straight line to the barycenters of the three subpopulations (red
line in panel a1 of Fig~\ref{f:histo}). Then, we rotated the
chromosome map in such a way that the fitted line is parallel to the
abscissa. The rotation point can be chosen arbitrarily, and we adopted
the center point of the chromosome map, at location (0.5, 0.5). We
computed the $\pm$1 $\sigma$ of the distribution of points along the
vertical axis of the rotated plane (corresponding to the pink lines in
panel a1), and constructed a histogram with all the stars within
$\pm$1$\sigma$ (in black in panel a2), and of each subpopulation again
within $\pm$1$\sigma$. The histogram of each subpopulation is
color-coded accordingly in panel (a2):\ bMS1 in blue, bMS2 in azure,
and bMS3 in cyan. In addition, we derived the histogram of a
distribution due to photometric and differential-reddening-correction
errors alone (in red in a2). The height of the error histogram has
been rescaled to be about half the height of the black histogram. This
is not an issue because what is important is the width of the error
histogram and not its height.

The histograms for MSd and MSe3+MSe4 stars were derived in the same
way as for bMS stars, and are shown in panels (b2) and (c2),
respectively.  Note that the first peak in the histogram distribution
of MSe3+MSe4 stars is due to the upper tail of the MSe2.

We can clearly see that the bMS3, the MSd3, and the MSe3 histogram
distributions have an extended tail to the right side of the
distribution, which might either indicate the presence of additional
subpopulations, some sort of systematic selection effects, or might
tell us that these subpopulations have experienced a prolonged period
of star formation, over which chemical abundances have changed
gradually.

We cannot exclude a-priori that what we have identified as the bMS2
(MSd3) is also due to an abundance spread within the bMS1 (MSd2)
caused by a prolonged star formation, rather than distinct peaks
associated with distinct subpopulation of the bMS (MSd). For the sake
of argument, in what follows we will keep considering the bMS2 and the
MSd3 as distinct subpopulations.

\begin{figure*}[!t]
\centering
\includegraphics[angle=-90,width=16cm]{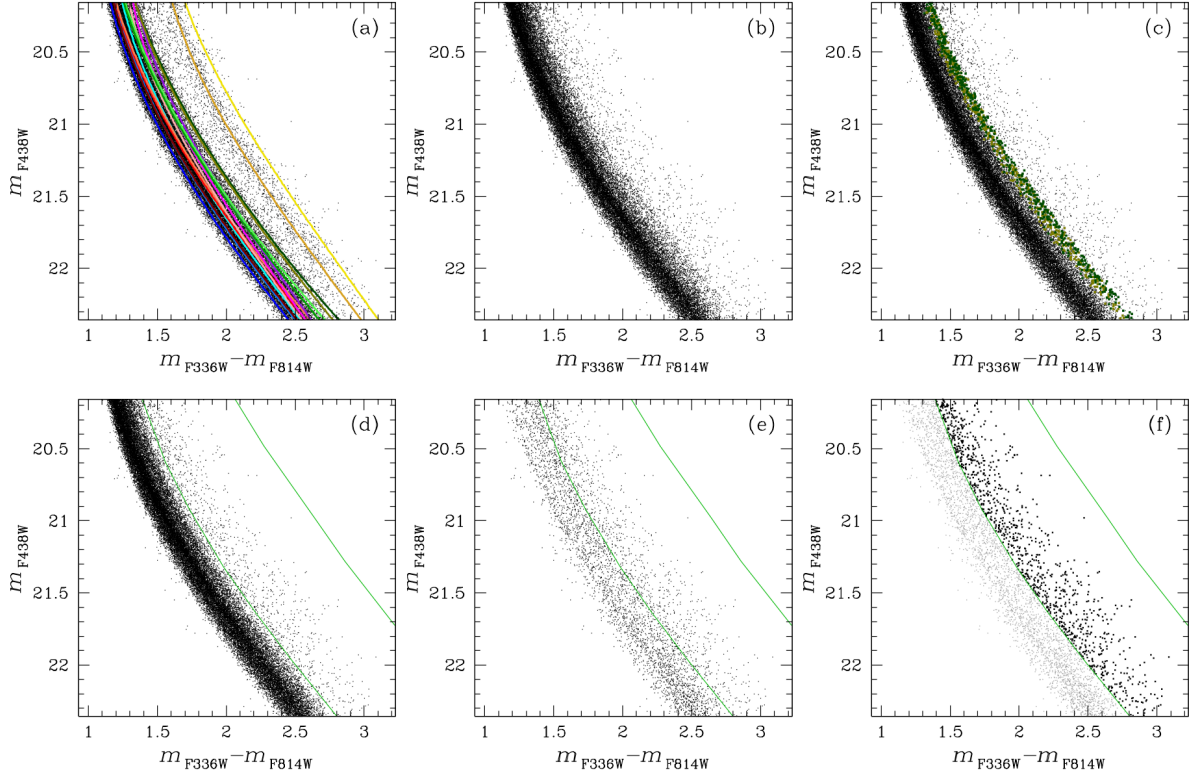}
\vskip -8mm
\caption{\small{(a) the $m_{\rm F336W}-m_{\rm F814W}$ CMD of
    \wcen\ between $20.35\le m_{\rm F438W}\le 22.55$. (b) same as (a),
    but we removed MSa1 and MSa2 stars. (c) same as (b), but we
    color-coded the redmost MSe3 and MSe4 subpopulations. (d) same as
    (c), but we also removed MSe3 and MSe4 stars. All other identified
    populations lie on the blue side of the left green line. The right
    green line shows the locus of MSa2-MSa2 equal-mass binaries, and
    no binary star can be redder than the right green line. (e) same
    as (b) but for unidentified stars only. We considered as binary
    stars all the unidentified stars between the two green
    lines. Binary stars are shown in black in panel (f), while the
    remaining unidentified stars are in gray.\\~\\}}
\label{f:bina}
\end{figure*}

\begin{figure*}[!t]
\centering
\includegraphics[angle=-90,width=16cm]{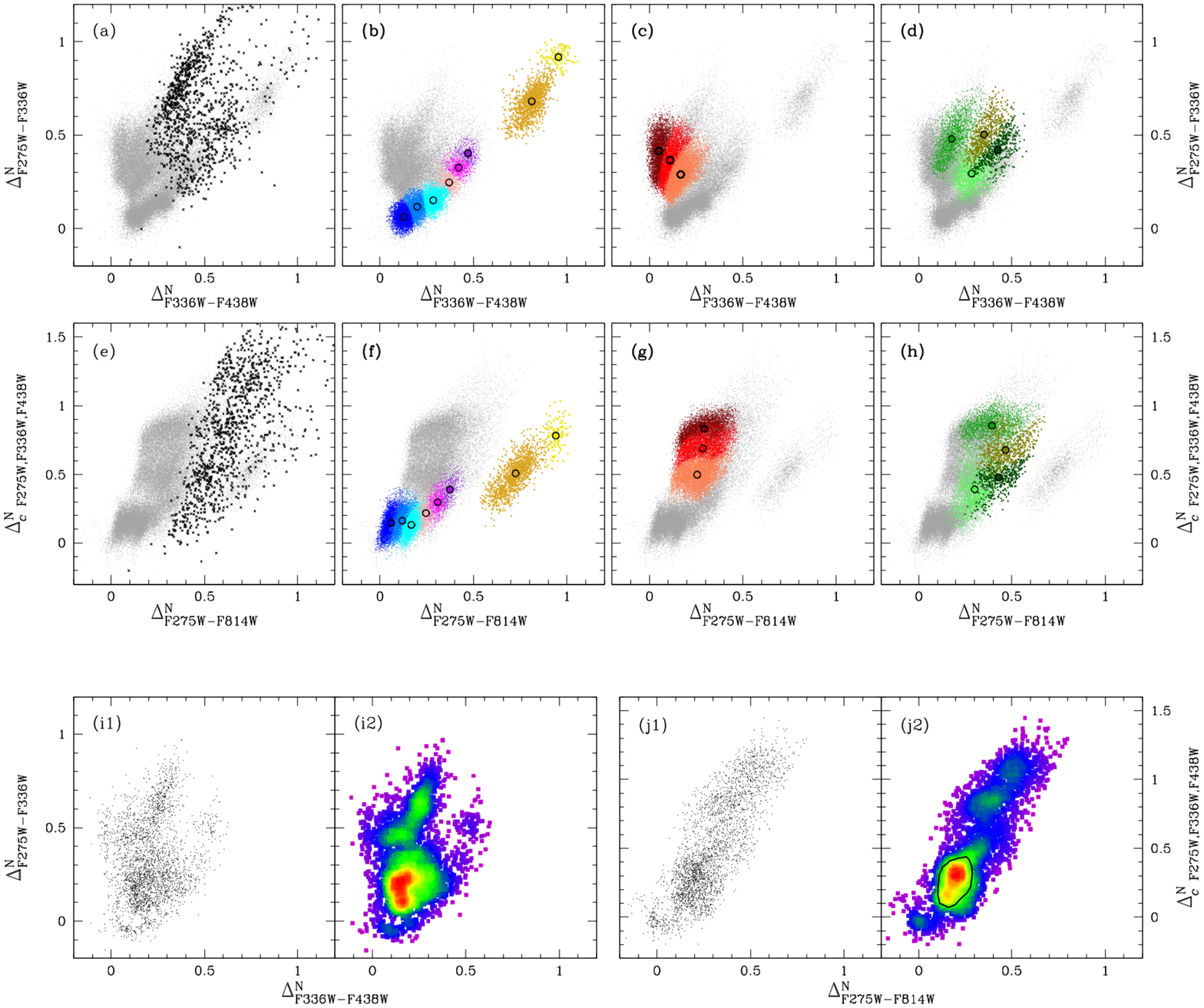}
\caption{The top panels show the $\Delta_{m_{\rm F275W}-m_{\rm
      F336W}}^{\rm N}$ vs.\ $\Delta_{m_{\rm F336W}-m_{\rm F438W}}^{\rm
    N}$ chromosome map of the entire MS of \wcen\ within $20.16\le
  m_{\rm F438W}\le 22.36$ . (a) MS single stars are in gray, while
  candidate binary stars are marked in black. (b) binary stars are
  removed, and MSa, bMS and MSd stars are color-coded according to
  their subgroups. The barycenter of each subpopulations is also
  marked with a solid circle with the corresponding color (as defined
  in the previous figures). (c) rMS stars are highlighted. (d) MSe
  stars are highlighted. The middle panels propose again the content
  of the top panels, but using a different chromosome
  map:\ $\Delta_{C\ {\rm F275W,F336W,F438W}}^{\rm N}$
  vs.\ $\Delta_{m_{\rm F275W}-m_{\rm F814W}}^{\rm N}$. Panels (i1) and
  (i2) show the $\Delta_{m_{\rm F275W}-m_{\rm F336W}}^{\rm N}$
  vs.\ $\Delta_{m_{\rm F336W}-m_{\rm F438W}}^{\rm N}$ chromosome map
  and its Hess diagram, respectively, for stars that do not belong
  neither to any of the previously-identifies subpopulations nor to
  the binary sample.  Panels (j1) and (j2) show the $\Delta_{C\ {\rm
      F275W,F336W,F438W}}$ vs.\ $\Delta_{m_{\rm F275W}-m_{\rm
      F814W}}^{\rm N}$ and its Hess diagram for the same
  previously-unidentified stars. The black boundary in panel (j2)
  encloses the main clump of stars. See the text for details.\\~\\}
\label{f:more}
\end{figure*}

\subsection{A qualitative estimate of the binary-fraction}
\label{ss:bina}

There are still 4087 unidentified MS stars (or $10.34\pm 0.17$\%) in
the magnitude interval $20.16\le m_{\rm F438W}\le 22.36$. First of
all, we wanted to identify and remove as many binaries as possible, so
that binaries will not bias any further attempt of identifying and
tagging additional subpopulations.  To this aim, we took advantage of
the fact that, with the exclusion of MSa stars, all other
\wcen\ subpopulations are almost overlapped to each other in the
$m_{\rm F336W}-m_{\rm F814W}$ CMD, which is illustrated in panel (a)
of Fig.~\ref{f:bina}. The fiducial line of the 15 subpopulations
identified so far are also shown, color-coded accordingly.  Panel (b)
is similar to panel (a), with the difference being that we removed the
fiducial lines and all stars belonging to the MSa1 and the MSa2. Of
the remaining 13 subpopulations, MSe3 and MSe4 stars are the reddest
ones and (slightly) isolated from the others. For clarity, in panel
(c) we show the same CMD as in panel (b) with just MSe3 and MSe4 stars
highlighted with their corresponding colors.

To better isolate likely binary stars, we also removed MSe3 and MSe4
stars and plotted the remaining stars in panel (d). It is clear from
the figure that the vast majority of the remaining 11 subpopulations
lie on the blue side of the left-most green line (drawn by hand). The
right-most green line, instead, defines the locus of MSa2+MSa2
equal-mass binaries. Since MSa2 stars are the reddest ones in this
CMD, we expect no binaries redward of the right-most green line
(besides, there are no stars redder that this line in the CMD).

Panel (e) of Fig.~\ref{f:bina} shows the same CMD region of the
previous panels, but this time we removed all stars belonging to the
15 previously-identified subpopulations. The two green lines in this
panel are the same as in panel (d).  Unidentified stars between the
two green lines are potential binaries (shown in black in panel f),
while stars bluer than the left-most green line (in gray in panel f)
are expected to be a combination of:\ (i) binaries, (ii) stars
belonging to already-identified subpopulations that have been somehow
rejected during our selection procedures, and --possibly but less
likely-- (iii) a few badly-measured sources that somehow made it into
our high-photometric-quality star list.

There are 1069 potential binaries ($2.70\pm 0.08$\%) in the magnitude
interval $20.16\le m_{\rm F438W}\le 22.36$ within the two green lines
(in black in panel f). This estimate could in principle represent a
lower limit, since more binaries can be present among the gray stars
in panel (f). On the other hand, among the 1069 potential binaries we
had identified could hide single stars with extreme chemical
compositions (more in the next Section). Our binary estimate is by no
mean intended to be rigorous, but it simply represents a convenient
tool that will allow us to further dissect the MS of the cluster.

\subsection{Are there more?}
\label{ss:more}

At this point, our TRSR iteration process is completed. No single CMD
(or pseudo-CMD) can further help us is unraveling additional
subpopulations. We can now have a look at where the 15 subpopulations
lie on chromosome maps derived using the entire MS.

The top and middle panels of Fig.~\ref{f:more} show the
$\Delta_{m_{\rm F275W}-m_{\rm F336W}}^{\rm N}$ vs.\ $\Delta_{m_{\rm
    F336W}-m_{\rm F438W}}^{\rm N}$ (top) and the $\Delta_{C\ {\rm
    F275W,F336W,F438W}}^{\rm N}$ vs.\ $\Delta_{m_{\rm F275W}-m_{\rm
    F814W}}^{\rm N}$ (middle) chromosome maps of the entire MS of the
cluster within the magnitude range $20.16\le m_{\rm F438W}\le 22.36$.
The chromosome-map plane of the top panels is the same we used in the
previous subsections. Note that, this time, the fiducial lines used to
rectify and parallelize the ${m_{\rm F275W}-m_{\rm F336W}}$ and
${m_{\rm F336W}-m_{\rm F438W}}$ CMDs had to necessarily enclose the
entire color extension of the MS and not just that of any single
population. As a consequence, the exact position of each star on the
top panels of Fig.~\ref{f:more} is different from that occupied by the
same star in the chromosome map of any previous figure. What matters
here is the \textit{relative} position of stars of different
subpopulations with respect to each other.

The chromosome map of the middle panels is instead similar to the one
extensively used by \citet{2017MNRAS.464.3636M} to analyze the
multiple populations on the RGB of 57 GCs (including that of
\wcen). \citet{2017MNRAS.464.3636M} used different fiducial lines to
rectify both the $m_{\rm F275W}-m_{\rm F814W}$ CMD and the $c_{\rm
  F275W,F336W,F438W}$ color index. Moreover, the convention we used
here for the $y$ axis of the middle panels in Fig.~\ref{f:more} is
opposite to that adopted by \citet{2017MNRAS.464.3636M}, so a more
direct comparison with the work of \citet{2017MNRAS.464.3636M} would
require an upside-down flip of our middle panels.  Again, what matters
here is the relative position occupied by stars of different
subpopulations.

In each of the top and middle panels, gray points represent MS single
stars. Selected binary stars are only shown in black in panels (a) and
(e), for reference. Let us briefly return to our binary selections. If
we compare the chromosome maps of Fig.~\ref{f:more} with that in
Fig.~6 of \citet{2017MNRAS.464.3636M} (but mind the flip of the
vertical axis) obtained for RGB stars, we can see that most of the
objects that are classified here as binaries would occupy the same
region as single stars with extreme [Fe/H], s-element, and C, N, and O
abundances of \citet{2017MNRAS.464.3636M}. The MS-based and the
RGB-based chromosome maps are similar but not identical, and while it
is possible that the different populations of \wcen\ would maintain
the same relative positions in the two planes, this cannot be taken as
a given. A deeper investigation, beyond the scope of the present
paper, is needed to solve this issue. For now, we stress that out
binary selections are not rigorous, and they should be taken ``cum
grano salis''. For simplicity, we will keep referring to these stars
simply as binaries.

In panels (b) and (f) of Fig.~\ref{f:more} we have highlighted bMS,
MSd and MSa subpopulations using the appropriate color-coding. The
barycenter of each subpopulation is also marked by a colored solid
circle. It is interesting to note that bMS, MSd and MSa stars appear
to be more or less aligned in panel (b), and are the bluest
subpopulations in $\Delta_{C\ {\rm F275W,F336W,F438W}}^{\rm N}$ at any
given value of $\Delta_{m_{\rm F275W}-m_{\rm F814W}}^{\rm N}$ (panel
f). These common features suggest that bMS, MSd and MSa stars must
share some similarities in their chemical composition.

The three rMS subpopulations are highlighted in panels (c) and (g),
while panels (d) and (h) show MSe stars. It is interesting to note
that rMS stars are aligned almost perpendicularly to the direction
defined by bMS, MSd and MSa stars in panel (c). Moreover, the three
rMS subpopulations are aligned almost vertically in panel (g).  It is
less obvious to find common features for MSe stars, other than the
fact that they occupy a well-defined region in both chromosome
maps. Nevertheless, if we were to tear the MSe apart and independently
consider the doublets (MSe1 + MSe2) and (MSe3 + MSe4), then it would
become apparent that both doublets share similar properties with those
of rMS stars. The two doublets are both aligned vertically in the
chromosome map of the middle panels and are both aligned
perpendicularly to the direction defined by bMS, MSd and MSa stars in
the chromosome map of the top panels.  It might make more sense to
actually separate the MSe subpopulations into two distinct groups,
with a new MSe formed by MSe1 and MSe2 stars, and --say-- a new MSf
formed by MSe3 and MSe4 stars. For now, however, such a separation
does not seem sufficiently justified, and we will keep considering the
MSe1, MSe2, MSe3 and MSe4 stars together within the same MSe group.

Finally, in the bottom panels of Fig.~\ref{f:more} we show the
chromosome maps and the companion Hess diagram for the remaining,
so-far unidentified 3018 stars. Panels (i1) and (i2) refer to the
$\Delta_{m_{\rm F275W}-m_{\rm F336W}}^{\rm N}$ vs.\ $\Delta_{m_{\rm
    F336W}-m_{\rm F438W}}^{\rm N}$ plane, while panels (j1) and (j2)
refer to the $\Delta_{C\ {\rm F275W,F336W,F438W}}^{\rm N}$
vs.\ $\Delta_{m_{\rm F275W}-m_{\rm F814W}}^{\rm N}$ plane.

The bottom panels reveal the presence of a prominent, well-defined
clump of stars, together with a few other lesser overdensities.  The
main clump, located at about (0.15,0.15) in panel (i2) and at about
(0.2,0.3) in panel (j2), is formed by the same stars in both
chromosome maps and peaks about half-way between the barycenters of
bMS and rMS stars.  Some of the other lesser overdensities in the
bottom panels might also be genuine new subpopulations of \wcen, but
there is a non-negligible chance that they might also be the
end-result of the conservative selection cuts we applied during the
population-tagging processes described in the previous
subsections. Moreover, some of these smaller clumps share the same
location of other subpopulations in one or both chromosome maps, or
they are suspiciously close to the regions occupied by binary stars to
make us think they might be binary stars as well.

The main clump is rounder and more compact in panel (j2), and possibly
exhibits a tail of points extending towards smaller $\Delta_{C\ {\rm
    F275W,F336W,F438W}}^{\rm N}$ values. Let us temporarily identify
the stars in this clump as MSx stars. At first, we thought the MSx
might constitute a new, genuine subpopulation of \wcen. We selected
all MSx stars within a the black boundary shown in panel (j2), and
plotted them in all the available CMDs. If the MSx were to be a
genuine subpopulation, then MSx stars would define a single sequence
in CMDs. It turned out that MSx stars appear to be split into two,
sometimes three segmented sequences in different CMDs.  It was clear
that the MSx is none other than a collection of stars belonging to
other subpopulations that were rejected preferentially in different
magnitude intervals. We will not discuss the specious MSx any
further. Since the main clump of unidentified stars in the chromosome
maps do not constitute an additional population of \wcen, the chances
for the lesser clumps to be themselves additional populations are
possibly even smaller than those for the MSx.

\section{Multiple-population overview}
\label{s:ov}

In the previous Section we described in detail how we were able to
photometrically isolate distinct MS populations (and their
subpopulations) in \wcen. The main observational findings can be
summarized as follows:\
\begin{itemize}
\item{We confirmed the findings of \citet{2010AJ....140..631B} of a
  split MSa;}
\item{We found that both the bMS and the rMS are each formed each by
  three subpopulations of stars (the rMS was previously shown to be
  split into two branches, see \citealt{2010AJ....140..631B});}
\item{We discovered a new population, the MSd, made up by three
  subpopulations, sharing similar properties to bMS and MSa stars;}
\item{We discovered another new population, the MSe, comprised of four
  subpopulations, sharing similar properties to rMS stars;}
\end{itemize}

\begin{figure*}[!t]
\centering
\includegraphics[angle=-90,width=18cm]{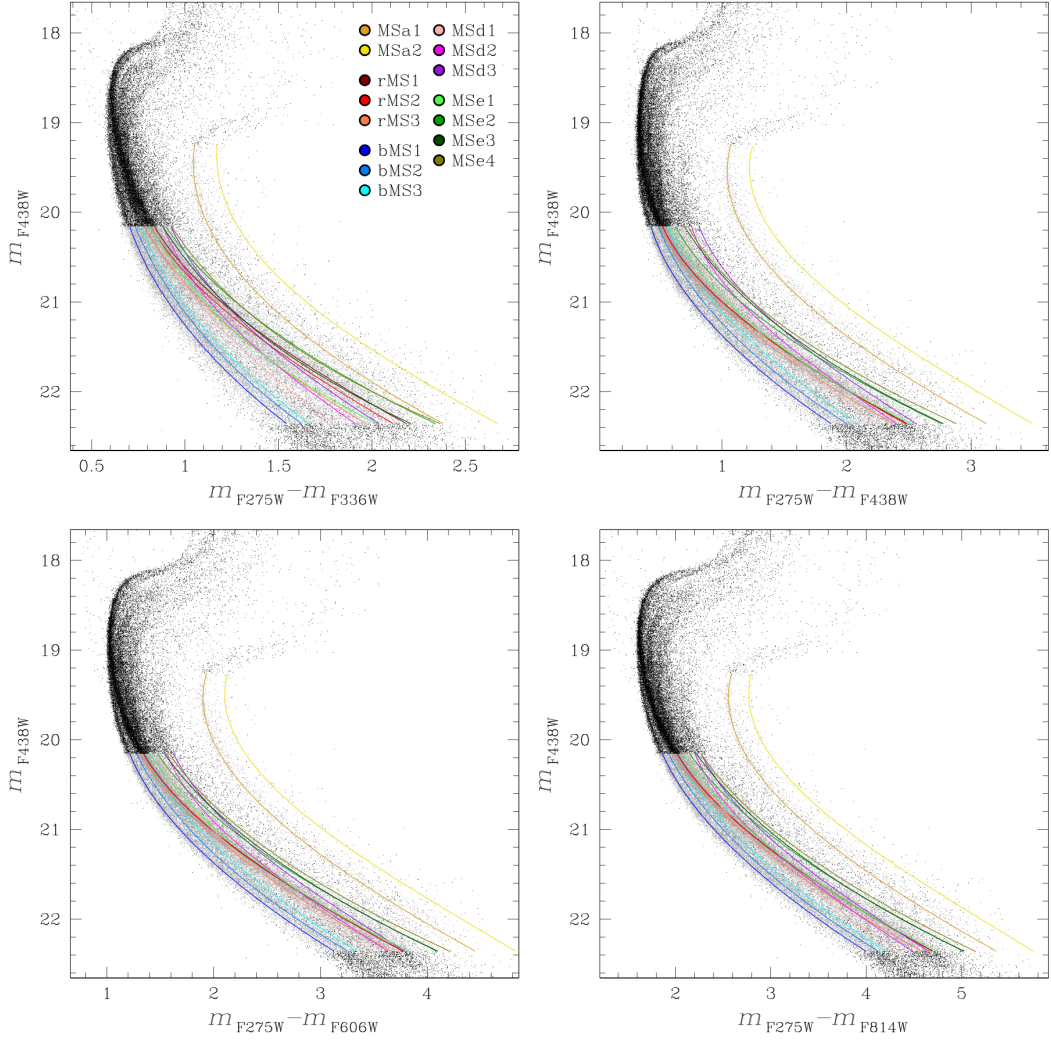}
\caption{\small{The $m_{\rm F275W}-m_{\rm F336W}$, $m_{\rm
      F275W}-m_{\rm F438W}$, $m_{\rm F275W}-m_{\rm F606W}$ and $m_{\rm
      F275W}-m_{\rm F814W}$ CMDs centered on the MS and SGB of \wcen.
    Identified stars are now in gray. Fiducial lines for each of the
    15 subpopulations are color-coded accordingly.\\~\\}}
\label{f:ov1}
\end{figure*}

\begin{figure*}[!t]
\centering
\includegraphics[angle=-90,width=18cm]{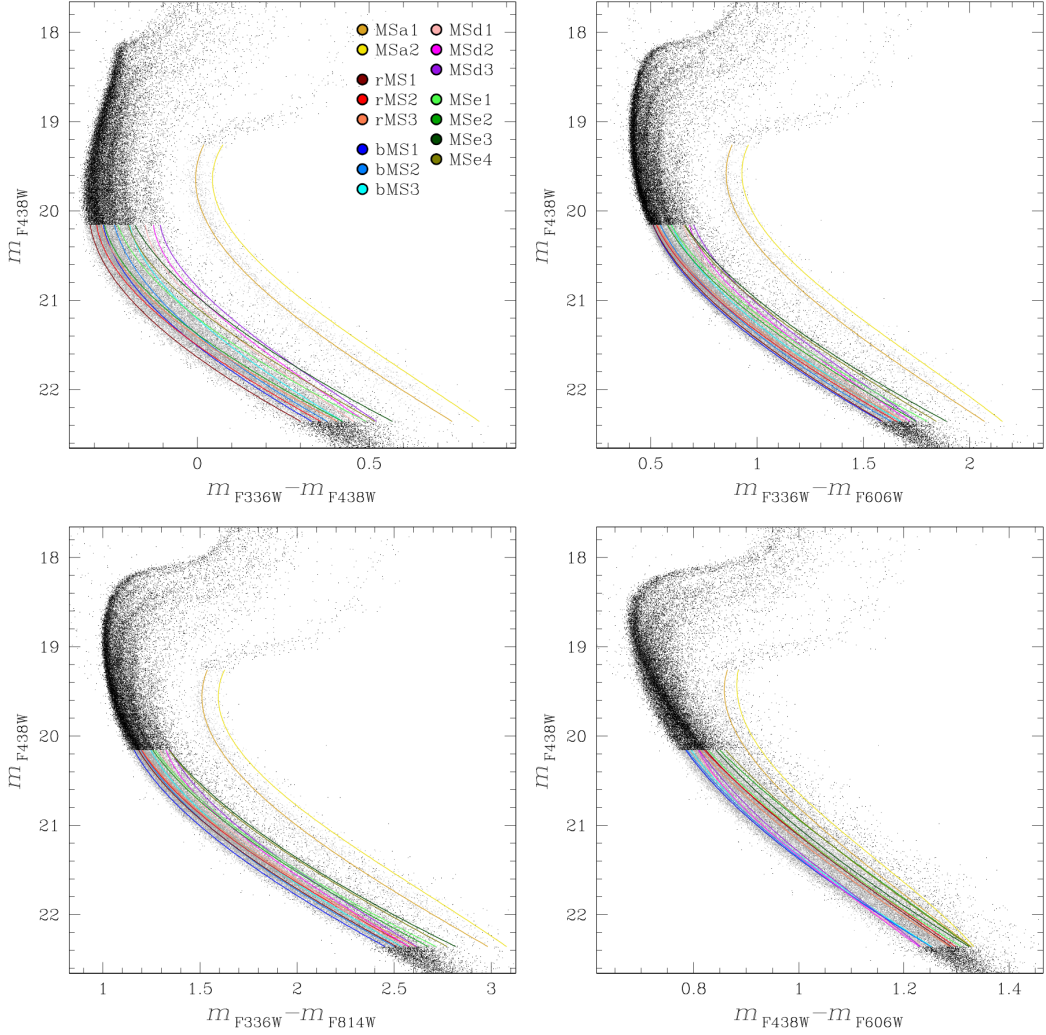}
\caption{\small{Similar to Fig.~\ref{f:ov1} but for the $m_{\rm
      F336W}-m_{\rm F438W}$, $m_{\rm F336W}-m_{\rm F606W}$, $m_{\rm
      F336W}-m_{\rm F814W}$ and $m_{\rm F438W}-m_{\rm F606W}$ CMDs.\\~\\}}
\label{f:ov2}
\end{figure*}

\begin{figure*}[!t]
\centering
\includegraphics[angle=-90,width=18cm]{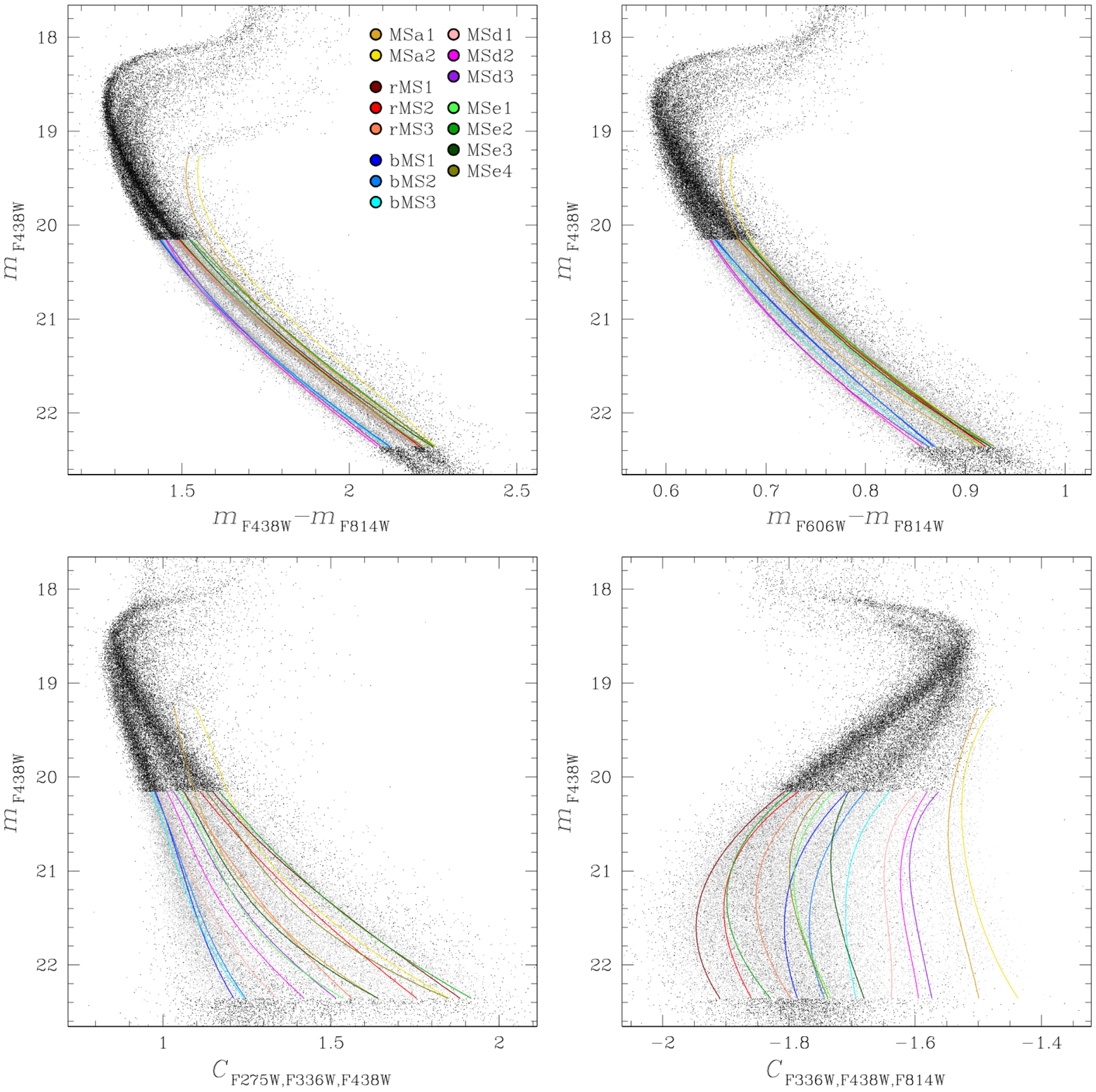}
\caption{\small{Similar to Fig.~\ref{f:ov1} but for the $m_{\rm
      F438W}-m_{\rm F814W}$ and $m_{\rm F606W}-m_{\rm F814W}$ CMDs and
    the two pseudo-CMDs $C_{\rm F275W,F336W,F438W}$ and $C_{\rm
      F336W,F438W,F814W}$.\\~\\}}
\label{f:ov3}
\end{figure*}

Figures~\ref{f:ov1}, \ref{f:ov2} and \ref{f:ov3} show all the possible
CMDs that can be obtained by combining the five filters at our
disposal (keeping $m_{\rm F438W}$ as the ordinate), zoomed-in around
the MS and SGB regions. For completeness, in Fig.~\ref{f:ov3} we also
included the $C_{\rm F275W,F336W,F438W}$ (bottom left) and the $C_{\rm
  F336W,F438W,F814W}$ (bottom right) pseudo-CMDs. The $C_{\rm
  F336W,F438W,F814W}$ pseudo-CMD represents the \textit{HST}
equivalent of the C$_{UBI}$ index (see, e.g.,
\citealt{2013MNRAS.431.2126M}).  We included in each panel of these
figures the fiducial line of the 15 subpopulations, color-coded
accordingly. For a better reading of the figures, identified stars are
now shown in gray.

We have based our population-tagging procedures on CMDs and chromosome
maps. Because of that, it was impossible to separate the different
subpopulations (besides the MSa group) for magnitudes brighter than
$m_{\rm F438W}$$\sim$20. If we had also made use of pseudo-CMDs (see,
e.g., the bottom panels of Fig.~\ref{f:ov3}), we might have been able
to push our selections one magnitude brighter. Including pseudo-CMDs
in our selection procedures would have added an extra layer of
complexity that goes beyond the scope of the present paper.  The
connection between the cluster's multiple populations at different
evolutionary stages will be the subject of a forthcoming paper. We
will use pseudo-CMDs to help us push the SGB selections to one
magnitude below the turn-off. This way, we will be able to directly
follow the 15 subpopulations from the MS up to the base of the RGB.

Some interesting subpopulation properties can be observed in these
CMDs:\
\begin{itemize}
\item{The MSa1 and MSa2 stars never cross each other. These two
  subpopulations are typically well separated from the others when the
  CMD color is based on the F275W or the F336W filters. On the other
  hand, MSa1 and MSa2 stars \textit{do overlap} with the rMS and the
  MSe in the $m_{\rm F606W}-m_{\rm F814W}$ CMD.}
\item{Of the three bMS subpopulations, the bMS1 is always the bluest
  one and the bMS3 always the reddest one when the CMD color is based
  on the F275W or the F336W filters, but the bMS3 becomes the bluest
  one in the $m_{\rm F606W}-m_{\rm F814W}$ CMD.  The bMS as a whole is
  typically the bluest population in all CMDs, with two notable
  exceptions:\ the rMS is the bluest one in the $m_{\rm F336W}-m_{\rm
    F438W}$ CMD, and the MSd is the bluest one in the $m_{\rm
    F606W}-m_{\rm F814W}$ CMD.}
\item{The rMS1 is the reddest of the three rMSs when the CMD color is
  of the form $m_{\rm F336W}-{\rm X}$, with X being any other redder
  filter, but the opposite happens in the $m_{\rm F275W}-m_{\rm
    F336W}$ CMD. Moreover, when the CMD color is of the form $m_{\rm
    F275W}-{\rm X}$, the three rMSs appear to curve towards redder
  colors at the faint end more than any other subpopulation (with the
  possible exception of the MSe3 and the MSe4).}
\item{The three MSd subpopulations (in particular the MSd2 and the
  MSd3) become increasingly more isolated the closer to the bright
  limit in the $m_{\rm F336W}-m_{\rm F438W}$ CMD, where they can
  easily be followed beyond the bright magnitude limit that we
  applied, all the way to the base of the RGB. It seems clear that the
  SGB-D group discussed in \citet{2014ApJ...791..107V} (called
  SGB4/MrI group by \citealt{2016MNRAS.457.4525T}) is the progeny of
  the MSd.}
\item{The three MSd subpopulations are as blue as the bMS in the
  $m_{\rm F438W}-X$ CMDs, but are the among the reddest populations
  (excluding MSa stars) in the $m_{\rm F336W}-m_{\rm F438W}$ CMD, and
  are generally overlapped with rMS and MSe stars in $m_{\rm
    F275W}-{\rm X}$ CMDs.}
\item{Stars of the MSe2 and the MSe4 are always redder than the
  three rMS subpopulations (but just marginally so in the $m_{\rm
    F606W}-m_{\rm F814W}$ CMD).}
\item{MSe1 and MSe3 stars are generally bluer than MSe2 and MSe4
  stars, but typically redder than rMS stars in most CMDs.}
\item{In the $C_{\rm F336W,F438W,F814W}$ pseudo-CMD, the MSe2 is
  generally overlapped with rMS stars, while MSe1, MSe3 and MSe4 stars
  overlap with bMS stars.}
\item{In the $C_{\rm F275W,F336W,F438W}$ pseudo-CMD, MSe2 stars are
  the reddest on average.}
\end{itemize}

\begin{figure*}[!t]
\centering
\includegraphics[angle=-90,width=\textwidth]{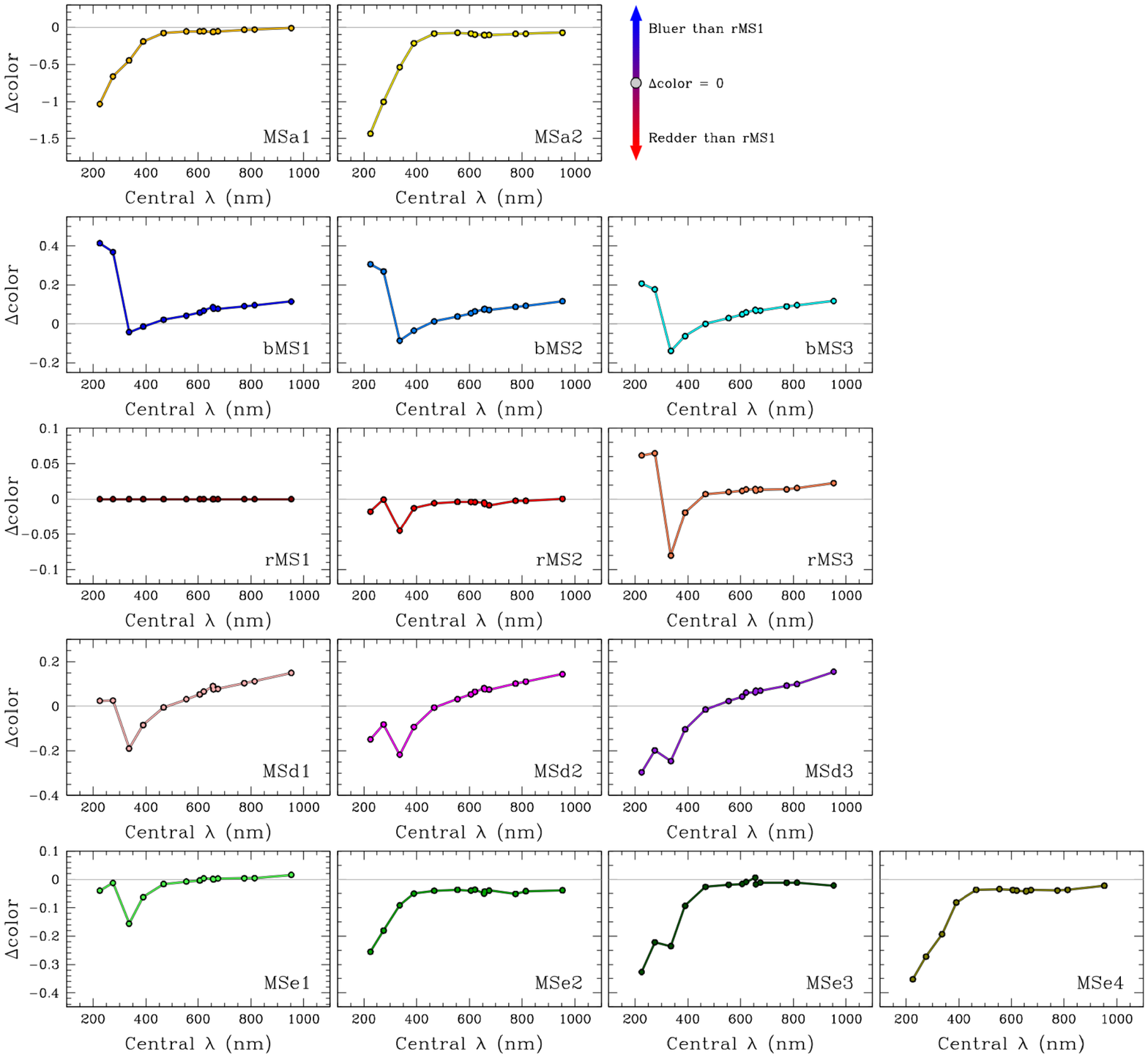}
\caption{\small{Color difference ($\Delta$color) between each
    subpopulation and the rMS1 (chosen as a reference) as measured in
    different CMDs of the form $X-m_{\rm F438W}$ or $m_{\rm F438W}-X$,
    where $X$ is any of filter other than F438W. The $x$ axis shows
    the central wavelength of the $X$ filter.  The $\Delta$color
    values are measured at the fixed magnitude level of $m_{\rm
      F438W}$=21.5.  Positive/negative $\Delta$color values imply that
    a given subpopulations is bluer/redder than the rMS1. Note that,
    in each panel, the vertical scale is kept fixed within each
    population group, but it varies for different groups. By
    definition, all $\Delta$color values of the reference
    subpopulation rMS1 are equal to zero, and are shown here only for
    completeness.\\~\\}}
\label{f:cd1}
\end{figure*}

\begin{figure*}[!t]
\centering
\includegraphics[angle=-90,width=16cm]{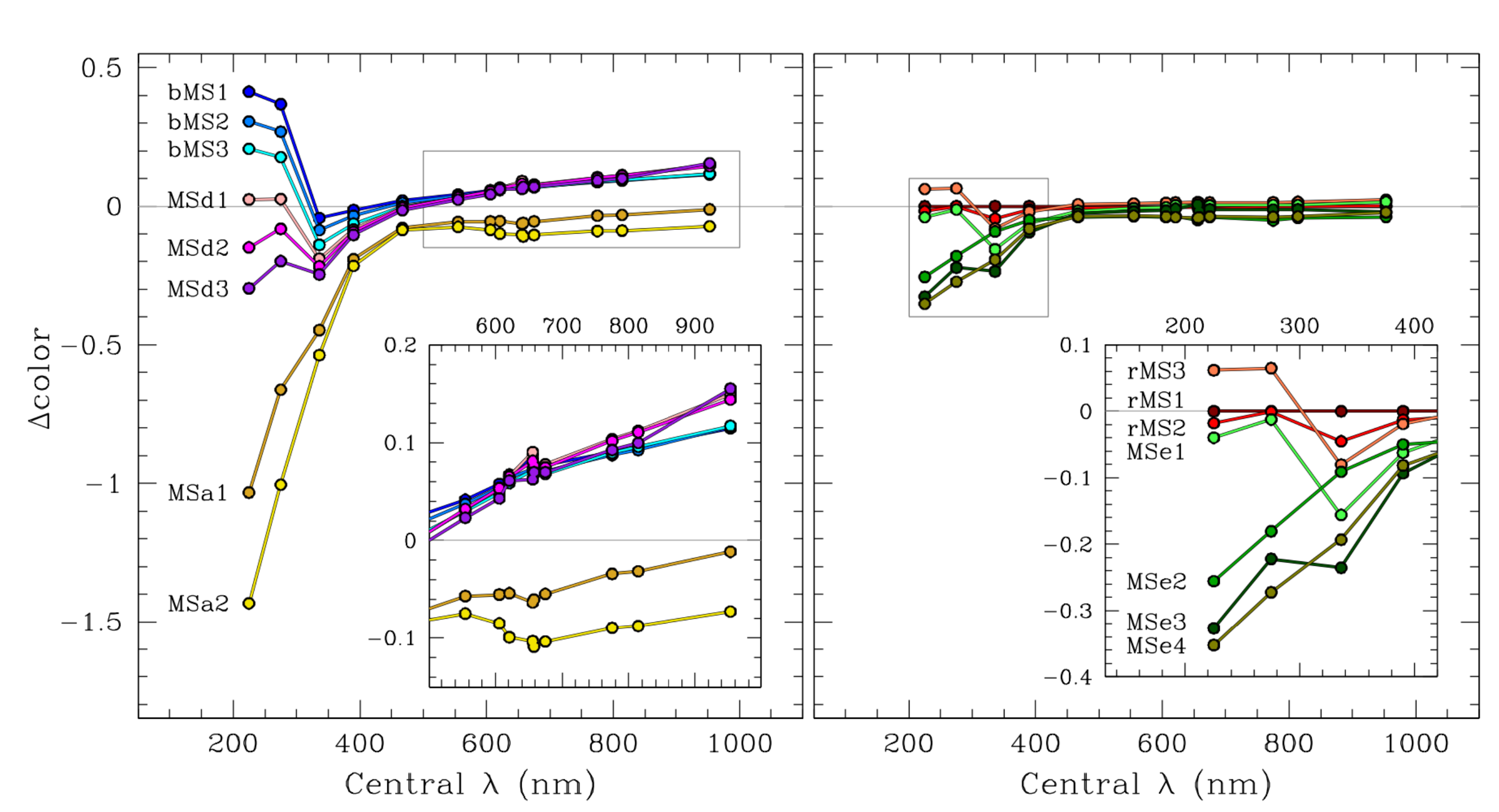}
\caption{\small{The $\Delta$color values of each subpopulations of
    Fig.~\ref{f:cd1} are collected in two groups:\ on the left those
    subpopulations sharing similar properties to the bMS, and on the
    right those subpopulations sharing similar properties to the
    rMS. On the left, all subpopulations show generally increasing
    $\Delta$color values for $\lambda\ge 336$ nm. Below $\lambda=336$
    nm, there is a gradually decreasing trend of the subpopulations,
    with the bMS1 reaching the highest $\Delta$color value at
    $\lambda=225$ nm, while the MSa2 reaches the lowest $\Delta$color
    value at the same $\lambda$. On the right, all subpopulations show
    a rather constant $\Delta$color value for $\lambda\ge 438$ nm, but
    then split into 2 groups (rMS1--3 + MSe1) and (MSe2--4) moving to
    shorter $\lambda$. The inset in each panel shows a zoomed-in view
    around some key characteristics of the each group.\\~\\}}
\label{f:cd2}
\end{figure*}

\begin{figure*}[t!]
\centering 
\includegraphics[width=16cm]{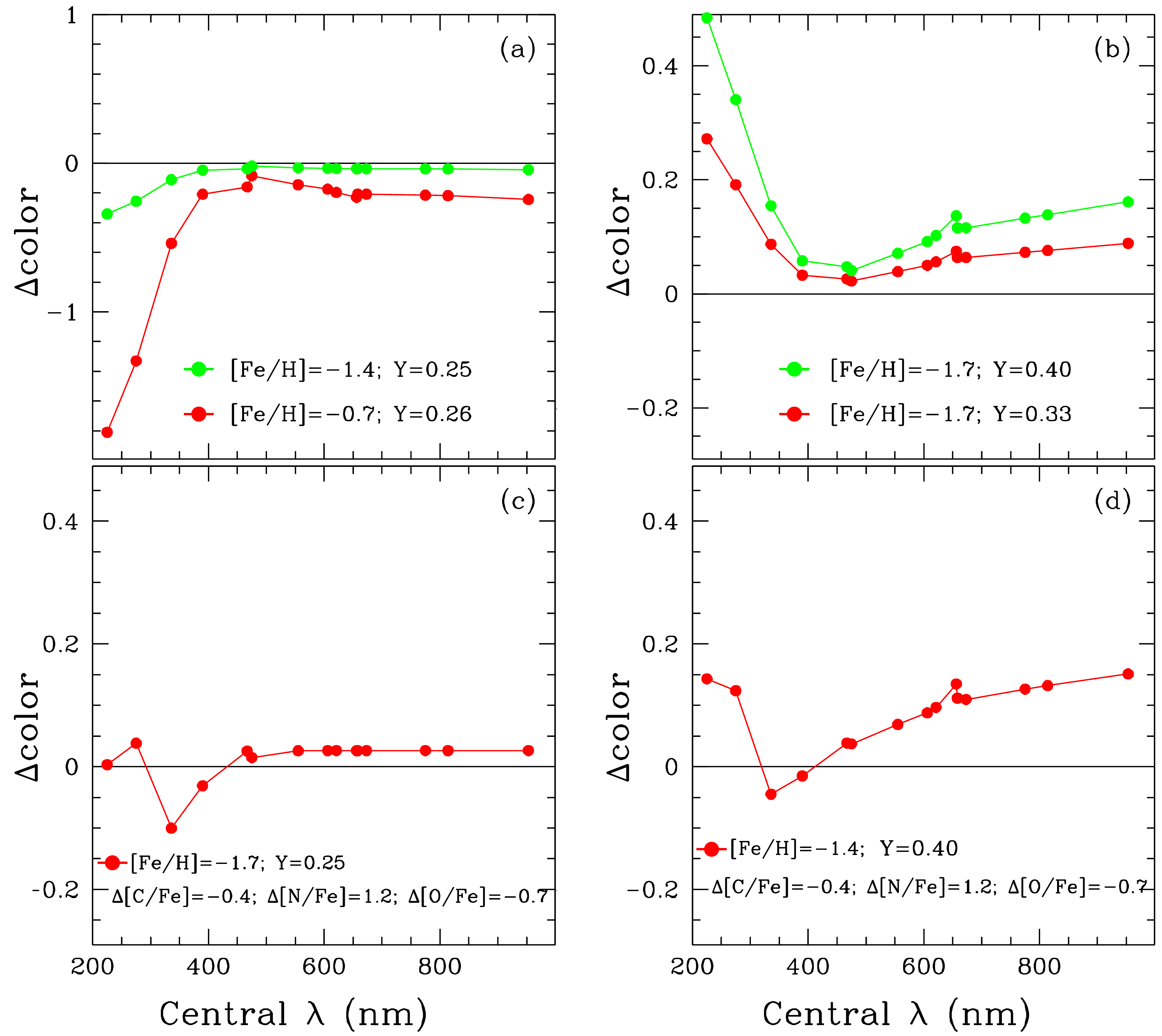}
\caption{\small{This figure shows the expected $\Delta$color versus
    wavelength behavior of stellar populations with different Fe, Y,
    and C, N, and O abundances with respect to a reference population
    with a chemical composition resembling that of the rMS1. Panel (a)
    shows two simulated populations with the same C, N, O and Y
    composition of the rMS1, but different [Fe/H] content
    ([Fe/H]=$-$1.4 is typical of bMS stars, and [Fe/H]=$-$0.7 is
    typical of MSa stars). In panel (b) we show the effects of
    different helium abundances with respect to the reference
    population, all other things being equal. The lower panels (c) and
    (d) illustrate the effects of changing Fe, He, C, N, and O, with
    results comparable to those of the two groups of populations with
    similar properties we collected in Fig.~\ref{f:cd2}. See the text
    for details.\\~\\}}
\label{f:th}
\end{figure*}

\section{Qualitative abundance analysis}
\label{s:ana}

In hopes of shedding more light onto this complicated observational
picture, we followed the powerful approach of looking at the color
separation of the 15 MS subpopulations as a function of the color
index in which they are observed (see, e.g.,
\citealt{2010AJ....140..631B,
  2013ApJ...765...32B,2012ApJ...744...58M,2013ApJ...767..120M,
  2015ApJ...808...51M}).  This analysis is also aimed at providing
some qualitative information for future interpretations of our
findings,

To do this, we measured the color difference ($\Delta$color) between
the rMS1, chosen as a reference, and each subpopulation.  The color
difference is measured in all CMDs of the form $X-m_{\rm F438W}$ or
$m_{\rm F438W}-X$ (depending on the filter $X$ being bluer or redder
than F438W, respectively) at the fixed magnitude level of $m_{\rm
  F438W}$=21.5. In this section, we extended the analysis to all the
UVIS filters at our disposal except for the the F350LP, F850LP (the
two long-pass filters) and the medium-band F390M (similar to the F390W
but with poorer photometry).  Note that we are keeping both F656N and
F658N filters. The former selects the H$_\alpha$ line, while the
latter selects the continuum after the H$_\alpha$ line. Finally, we
are not considering the case $X=m_{\rm F438W}$, since all the
corresponding $\Delta$color values are identical to zero (by
construction).

Figure~\ref{f:cd1} shows, for each subpopulation, the $\Delta$color
values as a function of the central wavelength (in nm) of the $X$
filter. Each row of panels in the figure refers to a different
population, from the MSa on the top to the MSe on the bottom, and are
organized in the same order we have selected them in
Sect.~\ref{s:mps}.  The panels within each row refer to the different
subpopulations of a given population group. The scale in the abscissa
is kept fixed in all panels, but the scale in the ordinate is only
kept fixed within a given population group, so that we can appreciate
both the small $\Delta$color differences between the rMS
subpopulations and the larger differences between the MSa
subpopulations and the reference rMS1. A gray horizontal line at
$\Delta$color=0 in each panel indicates no color differences with
respect to the rMS1 and is meant to better guide the eye.  We included
a colored arrow at the end of the top row to better read the direction
of the $\Delta$color.

We collected the $\Delta$color profiles of Fig.~\ref{f:cd1} into two
groups sharing similar properties. The first group contains those
subpopulations having increasing $\Delta$color values for $\lambda\ge
336$ nm, while the second group is made up by those subpopulations
with constant $\Delta$color values for $\lambda\ge 438$ nm. The bMS is
the archetype of the former group, and the rMS is the archetype of the
latter group. The $\Delta$color profiles of these two groups are shown
in Fig.~~\ref{f:cd2}. There is an interesting gradual
$\Delta$color-value decrease moving from the bMS1 to the MSa2 (left
panel). On the other hand, the subpopulations on the right panel seem
to split into 2 subgroups for $\lambda<300$ nm:\ the subgroup
(MSe2--4) has increasingly negative $\Delta$color values at shorter
wavelengths, while the subgroup (rMS1--3+MSe1) reaches a minimum
negative $\Delta$color at $\lambda=336$ nm before returning to
$\Delta$color values close to zero at shorter wavelength.

There are several stellar-abundance-variation effects dictating the
behavior of each profile we see in Figs.~\ref{f:cd1} and \ref{f:cd2}:\

\textit{C, N, O:} The F275W, F336W and F438W bandpasses include the
OH, HN, and CN+CH molecular bands, respectively. First-generation (1G)
stars are N-poor but O- and C-rich. As a result, all other things
being equal, 1G stars tend to be brighter in F336W and fainter in
F275W and F438W with respect to second-generation (2G) stars, which
are N-rich but C- and O-poor (see. e.g., \citealt{2013ApJ...767..120M}
for a detailed description and modeling of this phenomenon). As a
consequence, the higher the N abundance (and consequently the lower C
and O abundances), the deeper the negative $\Delta$color peak at
$\lambda$=336 nm, and the higher the rise at $\lambda$=225--275 nm.

\textit{Helium:} Optical filters are less sensitive to C, N, and O
variations, but offer a good proxy to temperature (and therefore
helium) variations. He-rich, 2G stars tend to have increasingly higher
$\Delta$color values at longer optical wavelengths the higher their He
content (\citealt{2013ApJ...767..120M}).

\textit{Iron:} The effects of different Fe abundances on the
$\Delta$color profiles are harder to quantify. To zeroth order, Fe and
He have similar effects on the stellar surface temperature. In
principle, the Fe-abundance could be estimated using the bluer part of
the profiles. In practice, the bluer part of the profiles is dominated
by light-element variations. Detailed Fe and C, N, and O abundances
have been published in the past for \wcen\ RGB stars (e.g.,
\citealt{2010ApJ...722.1373J, 2011ApJ...731...64M,
  2012ApJ...746...14M}) and SGB stars (e.g.,
\citealt{2014ApJ...791..107V}). A possible solution could be to link
each of the 15 MS subpopulations identified in this work with their
counterparts on the SGB and RGB, but this goes far beyond the purpose
of this paper, which is the detailed photometric analysis of the
multiple populations of \wcen\ along the MS. We plan to address this
topic in a forthcoming paper in this series.

In order to better understand the subpopulation trends we see in
Figs.~\ref{f:cd1} and \ref{f:cd2}, we compared the observed
$\Delta$color values with those predicted by appropriate synthetic
spectra (e.g., \citealt{2012ApJ...744...58M, 2013ApJ...767..120M,
  2014MNRAS.439.1588M, 2013ApJ...765...32B}).  The spectra are
computed for MS stars at a fixed magnitude of $m_{\rm F438W}$=21.5.

We assumed for the reference subpopulation rMS1 the following values:
[$\alpha$/Fe]=0.3, Y=0.25 (typical of a 1G population), [Fe/H]=$-$1.7,
[C/Fe]=0.4, [N/Fe]=0.2, [O/Fe]=0.3 (the latter three being typical
values for rMS stars, see., e.g., \citealt{2010ApJ...722.1373J,
  2011ApJ...731...64M, 2012ApJ...746...14M, 2017arXiv170400418M}).
Effective temperature and gravity values are based on the best-fitting
Dartmouth isochrones (\citealt{2008ApJS..178...89D}), obtained by
assuming a distance modulus (m$-$M)$_0$=13.69 and a reddening
E($B-V$)=0.13, similar to the values quoted in
\citet{1996AJ....112.1487H}, and an age of 13.5 Gyr.  The adopted
chemical and physical values were then used as input parameters for
the ATLAS12 and SYNTHE codes (\citealt{2005MSAIS...8...25C,
  2005MSAIS...8...14K, 2007IAUS..239...71S}) to generate a grid of
synthetic spectra in the wavelength range between 220 and 960
nm. These spectra were then integrated over the transmission curves of
the WFC3/UVIS filters used in this work.

Next, we synthesized stellar spectra for six populations characterized
by different [Fe/H], Y ,and C, N, O abundances. The results of these
simulations are summarized in Fig.~\ref{f:th}. Panel (a) shows how the
$\Delta$color varies as a function of wavelength if we only assume
different [Fe/H] abundances with respect to the reference
subpopulation. We adopted iron abundances typical of the bMS
([Fe/H]=$-$1.4) and of the MSa ([Fe/H]=$-$0.7).  In panel (b) we kept
the same iron abundance of the reference population, and we assumed
two different helium abundances that are typical of the extreme 2G
populations in \wcen\ (\citealt{ 2004ApJ...605L.125B,
  2004ApJ...612L..25N, 2005ApJ...621..777P}).  Panel (c) illustrates
the behavior of a population similar to the rMS1 in terms of iron and
helium, but with different C, N, and O abundances. Finally, in panel
(d) we show the effects of different C, N, and O abundances on a
population with primordial helium but [Fe/H] typical of the bMS.

By comparing the simulation results in Fig~\ref{f:th} to the
$\Delta$color trends of the 15 populations in Figs.~\ref{f:cd1} and
\ref{f:cd2}, we can qualitatively conclude the following:\
\begin{itemize}
\item{The subgroups of the bMS and the MSd are consistent with stellar
  populations highly enriched in both He and Fe with respect to the
  rMS1. Indeed, only a high He abundance can explain the increasing
  trends at longer optical wavelengths (panel b of Fig.~\ref{f:th}),
  making a sequence ---all other things being equal--- bluer in
  optical CMDs. (Variations in Fe and/or C, N, and O alone can only
  produce a flat trend at optical wavelengths, see panels a and c.)
  High Fe abundances must also be present to counter-balance/mitigate
  what would otherwise be a very steep increase of the trend at
  shorter wavelengths, also due to He. Since bMS stars are known to
  have a [Fe/H] of about $-1.4$, an even higher [Fe/H] value must be
  assumed for MSd stars, because of the less-positive trend at shorted
  wavelengths. The bMS and MSd groups also show evidence of N
  enrichment and C/O depletion, as inferred from the dip in the color
  profiles at 336 nm (i.e., using the F336W filter).}
\item{The rMS2, the rMS3 and the MSe1 should have a similar (low) Fe
  content to the rMS1, but different C, N, and O abundances. The rMS2
  and the MSe1 should also have a similar He content to that of the
  rMS1, while the rMS3 could slightly more Fe rich than the reference
  population (because of the decreasing trend at shorter
  wavelengths).}
\item{The observed pattern of the remaining three MSe subpopulations
  (MSe2, MSe3, and MSe4) seems mostly due to (small) iron variations,
  with possibly the MSe3 showing the signature of N differences.}
\item{The two MSa subpopulations are likely to be strongly enhanced in
  both iron and helium. As it is the case for bMS and MSd stars, only
  a high He abundance can explain the increasing trends at longer
  optical wavelengths. Moreover, a much higher Fe abundance than MSd
  stars must be assumed to completely reverse what would otherwise be
  an increasing, He-induced trend at shorter wavelengths. In fact,
  this is what we measure from spectra of RGB stars:\ the [Fe/H]
  abundance of the RGB equivalent of MSa stars (the so-called RGBa
  stars) is of about $-0.8$ (\citealt{2010ApJ...722.1373J,
    2012ApJ...746...14M}).}
\end{itemize}

\begin{figure*}[t!]
\centering 
\includegraphics[angle=-90,width=18cm]{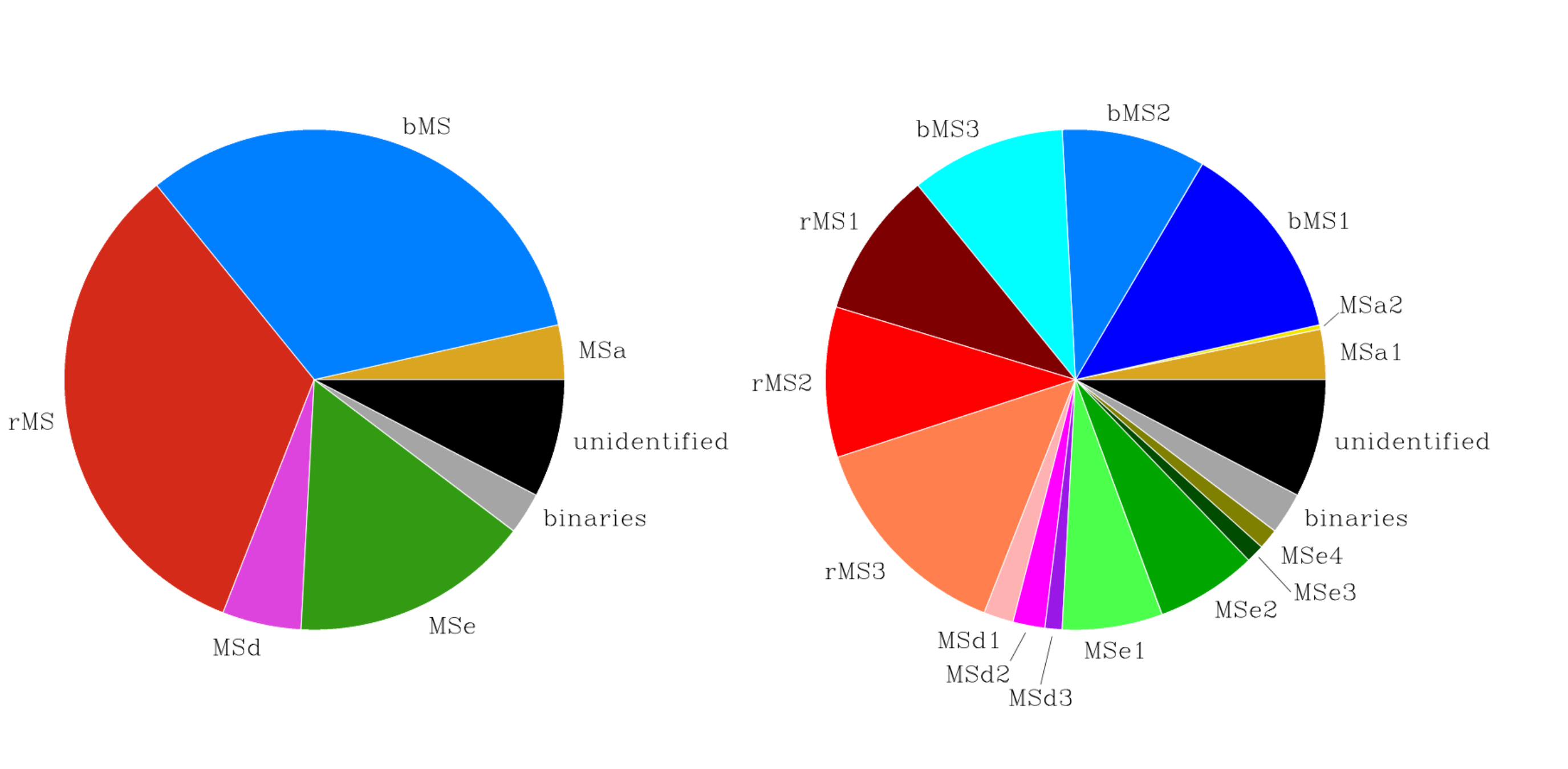}
\vskip -3mm
\caption{\small{Pie charts of the MS composition of \wcen\ in the
    magnitude interval $20.16\le m_{\rm F438W}\le 22.36$. The five
    main population groups are shown on the left panel, while their
    partition into subpopulations is on the right panel.\\~\\}}
\label{f:pie}
\end{figure*}

\begin{table}[t!]
\label{tab1}
\centering
\footnotesize{
\begin{tabular}{ccccccc}
\multicolumn{7}{c}{\textsc{Table~1}}\\
\multicolumn{7}{c}{\textsc{Multipopulation Components of the MS of \wcen}}\\
\multicolumn{7}{c}{\textsc{in the Magnitude Range $20.16<m_{\rm F438W}<22.36$}}\\
\hline\hline
$\!\!\!$\textsc{Main Group}$\!\!\!$&$\!\!\!$\textsc{Subgroup}$\!\!\!$&\textsc{N$_{\rm Stars}$}&\textsc{Fraction}&Fe&He&N\\
\hline
Entire MS&         & 39\,526 &   100\%&&&\\
\hline
MSa      &         & 1394    &  3.53$\pm$0.10\%&&&\\
         & MSa1    & 1283    &  3.25$\pm$0.09\%&+++&+++&?\\
         & MSa2    & 111     &  0.28$\pm$0.03\%&+++&+++&?\\
\hline
bMS      &         & 12\,776 & 32.32$\pm$0.33\%&&&\\
         & bMS1    & 5141    & 13.01$\pm$0.19\%&+&+++&++\\
         & bMS2    & 3683    &  9.32$\pm$0.16\%&+&+++&++\\
         & bMS3    & 3952    & 10.00$\pm$0.17\%&+&+++&++\\
\hline
rMS      &         & 13\,124 & 33.20$\pm$0.33\%&&&\\
         & rMS1    & 3739    &  9.46$\pm$0.16\%&$-$&$-$&$-$\\
         & rMS2    & 3838    &  9.71$\pm$0.16\%&$-$&$-$&+\\
         & rMS3    & 5547    & 14.03$\pm$0.20\%&$-$&$-$&+\\
\hline
MSd      &         & 2016    &  5.10$\pm$0.12\%&&&\\
         & MSd1    &  757    &  1.92$\pm$0.07\%&++&++&+\\
         & MSd2    &  819    &  2.07$\pm$0.07\%&++&++&+\\
         & MSd3    &  440    &  1.11$\pm$0.05\%&++&++&+\\
\hline
MSe      &         & 6129    & 15.51$\pm$0.21\%&&&\\
         & MSe1    & 2555    &  6.46$\pm$0.13\%&$-$&$-$&+\\
         & MSe2    & 2591    &  6.56$\pm$0.13\%&+&$-$&$-$\\
         & MSe3    &  463    &  1.17$\pm$0.05\%&+&$-$&+\\
         & MSe4    &  520    &  1.32$\pm$0.06\%&+&$-$&$-$\\
\hline
\hline
Binaries$^{(\dagger)}$ &         & 1069    &  2.70$\pm$0.08\%&&&\\
\hline
Unidentified &     & 3018    &  7.64$\pm$0.14\%&&&\\
\hline\hline
\multicolumn{4}{l}{($\dagger$) Included here for completeness.}\\
\end{tabular}}
\end{table}

\section{Conclusions}
\label{s:disc}

The exquisite photometric performance of \textit{HST}, in particular
in the near-UV regime, has allowed us to undertake for the first time
an extensive multi-color analysis of the MS of the GC \wcen. The main
results of this analysis can be summarized as follows. We have
confirmed the findings of \citet{2010AJ....140..631B} that the MSa is
split into two subpopulations (named here MSa1 and MSa2). Recently,
\citet{2017arXiv170400418M} identified six \wcen\ subpopulations along
the faint MS of the cluster, using exquisite \textit{HST} optical and
IR photometry of an external field (about 17$^\prime$ from the
cluster's center) from GO-9444, GO-10101 (both PI:\ King), GO-14118
and GO-14662 (both PI:\ Bedin). In particular, the authors found two
extreme subpopulations, which they named S1 and S2
(\citealt{2017arXiv170400418M}, their Fig.~2). Both S1 and S2
subpopulations are expected to be highly He-enhanced (Y$\sim$0.4), and
have an iron abundance [Fe/H] of $-$1.1 and $-$0.7,
respectively. Given the high iron abundance of the latter, it is
likely that the subpopulation S2 of \citet{2017arXiv170400418M} is
associated with the MSa.

We discovered that both the bMS and the rMS are actually split into
three subcomponents (the rMS was previously known to be split into two
subpopulations, \citealt{2010AJ....140..631B}). Moreover, we
discovered two additional population groups:\ the MSd and the MSe. The
former is itself split into three subcomponents, sharing properties
more similar to those of the bMS and the MSa. The latter is split into
four subcomponents, with properties more similar to the rMS.  Some of
these subpopulations also show hints of further subdivisions.  While
it is tempting to link the four main population groups rMS, MSe, MSd
and bMS to populations A, B, C, and D of \citet{2017arXiv170400418M},
respectively, it is worth noting that the bMS and the rMS have a
strong radial gradient (see, e.g., \citealt{2007ApJ...654..915S,
  2009A&A...507.1393B}). A clear association between the multiple
subpopulations we have found in the core and those found by
\citet{2017arXiv170400418M} at 17$^{\prime\prime}$ from the center of
the cluster would require either a chemical tagging or a
radial-gradient analysis of the relative population ratios.

In Sect.~\ref{s:mps}, we listed the relative number of stars in each
subpopulation and their overall fraction with respect to the analyzed
MS stars, but these numbers are reported in the text of the various
subsections. For convenience, we have collected these pieces of
information in Table~1. The quoted errors refer to Poisson errors
only. Fractions are rounded off to the closest hundredth.  It stands
to reason that slightly different selections in the chromosome maps of
each population would have led to (slightly) different subpopulation
fractions. The main purpose of our selections is to allow us to infer
qualitative properties of each subpopulation,; they are not meant to
be taken in an absolute sense. The quoted fractions for the five main
population groups are more likely reliable to about 10\%. The last
three columns of Table~1 qualitatively show the iron, helium, and
nitrogen relative abundances, respectively, with respect to the
reference subpopulation rMS1, based on the comparison with synthetic
spectra. One or more ``$+$'' signs indicate an increasingly high
abundance, a ``$-$'' sign refers a relative abundance similar to that
of the reference subpopulation, and a ``?''  means that that abundance
cannot be qualitatively quantified with the tools at our disposal.
The left panel of Fig.~\ref{f:pie} shows a pie chart of the main
population groups we found on the MS. In the right panel we further
divided each population into its own subcomponents.

We have based our population-tagging procedures on CMDs and chromosome
maps. Because of that, it would be hard to separate the different
subpopulations (besides the MSa group) for magnitudes brighter than
$m_{\rm F438W}$=20.16. If we had also made use of pseudo-CMDs (see,
e.g., the bottom panels of Fig.~\ref{f:ov3}, we might have been able
to push our selections one magnitude brighter. The connection between
the cluster's multiple populations on different evolutionary stages
will be the subject of a forthcoming paper. We will use pseudo-CMDs to
help us push the SGB selections to one magnitude below the
turn-off. This way, we will be able to directly follow the 15
subpopulations from the MS up to the RGB.

We make publicly available with this paper a two-column catalog with
the adopted selections.  The file contains 478\,477 lines, one for
each of the stars listed in the astro-photometric catalogs we
published in Paper~I. The first column contains the stellar ID numbers
as defined in Paper~I.  Values in the second column allow the user to
select stars belonging to the 15 subpopulations we have identified as
follows: 1=MSa1, 2=MSa2, 3=bMS1, 4=bMS2, 5=bMS3, 6=rMS1, 7=rMS2,
8=rMS3, 9=MSd1, 10=MSd2, 11=MSd3, 12=MSe1, 13=MSe2, 14=MSe3, and
15=MSe4. A value of ``0'' refers to any other star in the
astro-photometric catalog, including those that did not qualify for
this analysis. An extract of the companion catalog is shown in
Table~2.

\begin{table}[t!]
\label{tab2}
\centering
\begin{tabular}{rr}
\multicolumn{2}{c}{\textsc{Table~2}}\\
\multicolumn{2}{c}{\textsc{Multipopulation Identifier}}\\
\hline\hline
Star ID&Population ID\\
\hline
1&0\\
2&0\\
3&0\\
4&0\\
5&0\\
\dots & \dots \\
17042& 0\\
17043& 5\\
17044& 0\\
17045& 7\\
17046& 0\\
\dots & \dots \\
\hline\hline
\end{tabular}
\end{table}

\acknowledgments \noindent \textbf{Acknowledgments.} AB acknowledges
support from STScI grants AR-12656 and AR-12845.  APM and AFM
acknowledge support by the Australian Research Council through
Discovery Early Career Researcher Awards DE150101816 and DE160100851.
GP acknowledges partial support by PRIN-INAF 2014 and by the "Progetto
di Ateneo 2014 CPDA141214 by Universit\`a di Padova.

\small{
}
\end{document}